\documentclass[11pt, a4paper]{article}
\usepackage[a4paper, margin=2.5cm]{geometry}
\usepackage[T1]{fontenc}
\usepackage{lmodern}
\usepackage[english]{babel}
\usepackage{amsmath, amssymb, amsthm}
\usepackage{booktabs}
\usepackage{enumitem}
\usepackage[hidelinks]{hyperref}

\newtheorem{theorem}{Theorem}[section]
\newtheorem{lemma}[theorem]{Lemma}
\newtheorem{proposition}[theorem]{Proposition}

\theoremstyle{definition}
\newtheorem{definition}[theorem]{Definition}
\theoremstyle{remark}
\newtheorem{remark}[theorem]{Remark}

\numberwithin{equation}{section}
\allowdisplaybreaks[2]
\setlist{itemsep=3pt, topsep=5pt, parsep=0pt}

\title{An $O(n\log n)$ Algorithm for Uncapacitated Single-Item Lot Sizing with a One-Breakpoint All-Units Discount and Arbitrary Prices}
\author{Kleitos Papadopoulos\thanks{Email: \texttt{kleitospa@gmail.com}}}
\date{}

\begin{document}
\maketitle

\begin{abstract}
We study the finite-horizon, uncapacitated single-item lot-sizing problem with no fixed setup cost, a stationary threshold $Q > 0$ for an all-units quantity discount, arbitrary time-varying regular and discounted unit prices, and nonnegative linear holding costs. Order quantities and inventories may be either continuous or integral, and neither backlogging nor free disposal is allowed. Because unit prices may rise and fall arbitrarily over time, an optimal physical schedule may engage in speculative forward buying: positive orders can occur when incoming stock already exceeds current demand, and physical inventory carried across periods can grow proportionally to total horizon demand rather than remaining bounded by a multiple of $Q$.

We give an exact $O(n\log n)$-time algorithm that computes the minimum total cost with free terminal inventory, together with the optimal cost for every horizon prefix constrained to end with zero inventory, using $O(n)$ working space under unit-cost exact arithmetic. Value computation operates online in a single forward pass without lookahead. With persistent frontier snapshots and a linear-time retroactive order-aggregation pass, an optimal physical ordering schedule is recovered in $O(n\log n)$ time and $O(n\log n)$ space.

The proof combines three ingredients. First, tracking running minima of holding-cost-adjusted historical prices reduces the arbitrary-price problem online to an exact virtual lot-sizing problem whose effective prices satisfy non-speculative temporal inequalities, paired with a tier-preserving origin map back to physical orders. Second, minimizing a linear inventory potential over optimal virtual schedules establishes a simultaneous normal form under non-speculative prices: virtual orders occur only when incoming virtual stock is strictly below current demand, intermediate virtual stock remains strictly below $2Q$, and free-terminal virtual stock ends strictly below $Q$. This justifies a bounded-domain auxiliary dynamic program whose normalized high-inventory envelope exhibits a single crossover despite downward discount jumps. Third, an augmented balanced search tree over ordered one-kink candidate curves maintains the auxiliary frontier via an exact three-regime comparator threshold, lazy bulk endpoint snapshots, and guarded single-crossing merges in $O(n)$ total structural events.
\end{abstract}


\section{Introduction}
\label{sec:introduction}

Single-item dynamic lot sizing is a fundamental problem in inventory and production planning: given deterministic demands $d_1, \dots, d_n$ over a finite planning horizon of $n$ periods, one chooses nonnegative order quantities $q_1, \dots, q_n$ to satisfy demand without backlogging while minimizing total procurement and inventory holding costs. Under an \emph{all-units} quantity discount with threshold $Q > 0$, ordering $0 \le q_i < Q$ units in period $i$ incurs a regular unit price $p_{1,i}$, whereas ordering $q_i \ge Q$ units lowers the unit price on the \emph{entire} batch to $p_{2,i} \le p_{1,i}$. Because the procurement cost function drops discontinuously at $q_i = Q$, purchasing surplus stock to reach the threshold can be strictly cheaper than ordering the exact net demand, even when holding costs are positive and disposal of excess inventory is prohibited.

When both price sequences $\{p_{1,i}\}$ and $\{p_{2,i}\}$ are restricted to be non-increasing over time, recent work~\cite{papadopoulos2026monotone} showed that single-item lot sizing with a stationary one-breakpoint all-units discount and zero setup cost can be solved in $O(n\log n)$ time, even under time-varying post-order capacities. Once prices are allowed to vary \emph{arbitrarily} over time, however, the structural properties that directly bound physical inventory break down. Ahead of a sharp price increase, an optimal physical schedule may purchase large speculative batches even when incoming inventory already covers several upcoming periods of demand, carrying stock arbitrarily larger than $2Q$. Conversely, when prices subsequently drop, previously accumulated stock competes with newly discounted batches across nonconvex jump discontinuities.

\subsection{Main results and proof architecture}

In this paper we resolve the uncapacitated problem under completely arbitrary nonnegative unit prices $0 \le p_{2,i} \le p_{1,i}$ and nonnegative linear holding costs $h_i \ge 0$, for both continuous ($\mathcal{G} = \mathbb{R}_{\ge 0}$) and integral ($\mathcal{G} = \mathbb{Z}_{\ge 0}$) quantity domains. The formal model and arithmetic conventions are stated in Section~\ref{sec:model}.

\begin{theorem}[Optimal values in $O(n\log n)$ time and linear space]
\label{thm:main}
For any $n$-period uncapacitated instance satisfying \eqref{eq:price-tier}--\eqref{eq:feasibility}, the minimum total cost $\operatorname{OPT}$ with free terminal inventory, together with the zero-terminal prefix optimum $\operatorname{OPT}_{k,0}$ for every $k \in \{1, \dots, n\}$, can be computed online in $O(n\log n)$ time and $O(n)$ working space. These bounds hold for both $\mathcal{G} = \mathbb{R}_{\ge 0}$ and $\mathcal{G} = \mathbb{Z}_{\ge 0}$ under the unit-cost exact arithmetic model.
\end{theorem}

\begin{theorem}[Optimal physical schedule recovery]
\label{thm:recovery-main}
Under the hypotheses of Theorem~\ref{thm:main}, an optimal physical order sequence $(q_1, \dots, q_n) \in \mathcal{G}^n$ for the free-terminal problem---or for any prescribed zero-terminal prefix $1, \dots, k$---can be computed in $O(n\log n)$ time and $O(n\log n)$ space.
\end{theorem}

The proof is organized around three main steps that decouple speculative physical stockpiling from compressed frontier maintenance:

\begin{enumerate}
\item \textbf{Online effective-price reduction and retroactive aggregation (Section~\ref{sec:effective-reduction}).}
At each period $i$, we maintain in $O(1)$ time the cheapest holding-cost-adjusted historical regular price $\alpha_i$ and discounted price $\beta_i$, together with their attaining origin periods $s(i), r(i) \le i$. Replacing $(p_{1,i}, p_{2,i})$ by $(\alpha_i, \beta_i)$ defines a \emph{virtual} lot-sizing problem whose prices are no longer arbitrary: they satisfy the \emph{non-speculative} inequalities $\alpha_{i+1} \le \alpha_i + h_i$ and $\beta_{i+1} \le \beta_i + h_i$. Theorem~\ref{thm:effective-equivalence} proves that the virtual and physical problems have identical free-terminal and zero-terminal optima on every horizon prefix. Crucially, routing virtual orders back to their origin periods $(s(i), r(i))$ and aggregating colliding orders never worsens price tiers: combining regular orders that cross $Q$ only unlocks a cheaper discounted rate $p_{2,t} \le p_{1,t}$, while discounted virtual orders already have size at least $Q$.

\item \textbf{Simultaneous normal form under non-speculative prices (Sections~\ref{sec:normal-form}--\ref{sec:auxiliary-dp}).}
Unlike monotone prices ($\beta_{i+1} \le \beta_i$), non-speculative effective prices permit $\alpha_{i+1} > \alpha_i$ and $\beta_{i+1} > \beta_i$ when $h_i > 0$. By accounting jointly for procurement and cumulative holding savings during tier-preserving history trimming, we prove that minimizing the virtual inventory potential $\Phi(\hat{I}) = \sum_{t=1}^m \hat{I}_t$ yields one optimal virtual schedule that simultaneously satisfies $\hat{q}_i > 0 \implies \hat{I}_{i-1} < d_i$, $\hat{I}_{i-1} < 2Q$, and $\hat{I}_n < Q$. This justifies an auxiliary dynamic program supported on $[0, 2Q]$ that overestimates certain positive-inventory states without altering $\operatorname{OPT}$ or any $\operatorname{OPT}_{k,0}$. Subtracting $\beta_i y$ makes the regular departure envelope nondecreasing above $Q$ and the discounted envelope nonincreasing, establishing a single takeover crossover even across downward discount jumps.

\item \textbf{Certified one-kink frontier and amortized balanced tree (Sections~\ref{sec:representation}--\ref{sec:recovery}).}
We represent the auxiliary frontier as the lower envelope of interleaved one-kink candidate curves stored in an augmented height-balanced search tree. Adjacent dominance under regular-price updates is governed by an exact three-regime signed comparator threshold $\theta_j$ that translates uniformly by $+h_i$ under holding shifts. Bulk discounted transitions reduce to prefix-record minima of parallel rays normalized lazily via frozen successor-anchor snapshots, and merges locate the unique takeover coordinate in $O(\log n)$ time via a bracket-guarded binary search. Every nonterminal loop step permanently deletes a candidate node, bounding the total number of structural tree events by $O(n)$. Two-stage recovery combines path-copying persistence on the auxiliary frontier with $O(n)$ origin aggregation.
\end{enumerate}

\subsection{Relation to prior work}

Dynamic lot sizing originates with Wagner and Whitin~\cite{wagner1958dynamic}, who gave an $O(n^2)$ forward algorithm for uncapacitated lot sizing with fixed setup costs and stationary or monotone unit costs. Nearly three decades later, Wagelmans, van Hoesel, and Kolen~\cite{wagelmans1992economic} and Aggarwal and Park~\cite{aggarwal1993improved} independently improved the classical uncapacitated Wagner--Whitin problem with arbitrary time-varying unit production costs to $O(n\log n)$ time via geometric envelope maintenance and Monge-array techniques; see also Chung and Lin~\cite{chung1988algorithm} and Atamt\"urk and Hochbaum~\cite{atamturk2001capacity} for polynomial algorithms in structured capacitated settings.

Lot sizing under quantity discounts introduces a fundamentally different nonconvexity because the cost break occurs at a positive threshold $Q > 0$ rather than at zero. Federgruen and Lee~\cite{federgruen1990dynamic} analyzed uncapacitated dynamic lot sizing with setup costs and stationary or non-increasing cost parameters, obtaining an $O(n^2)$ algorithm for incremental discounts and an $O(n^3)$ algorithm for a one-breakpoint all-units discount. Extensions incorporating resale or capacitated setup structures require higher-degree polynomial bounds, such as $O(n^4)$ in Malekian, Mirmohammadi, and Bijari~\cite{malekian2021polynomial}.

For uncapacitated lot sizing with general time-varying piecewise-linear or piecewise-concave procurement costs and $m$ stationary breakpoints, Koca, Yaman, and Akt\"urk~\cite{koca2014lot} established polynomial solvability in $O(n^{2m+3})$ time ($O(n^5)$ for $m = 1$). Ou~\cite{ou2017improved} improved the piecewise-linear bound to $O(n^{m+2}\log n)$ ($O(n^3\log n)$ for $m = 1$) using range-minimum queries, and Papadopoulos~\cite{papadopoulos2026piecewise} removed the logarithmic factor to obtain $O(n^{m+2})$ ($O(n^3)$ for $m = 1$). Those general formulations allow nonzero fixed setup intercepts on each segment. When setup costs are absent and the nonconvexity stems purely from a one-breakpoint all-units discount, Papadopoulos~\cite{papadopoulos2026monotone} achieved $O(n\log n)$ time under non-increasing unit prices. Table~\ref{tab:literature} summarizes representative exact complexity bounds; the present work shows that in the uncapacitated setting without setup costs, removing price monotonicity entirely incurs no asymptotic overhead beyond $O(n\log n)$.

\begin{table}[tbp]
\centering
\small
\begin{tabular}{@{}p{3.6cm}p{4.6cm}p{3.4cm}p{2.2cm}@{}}
\toprule
\textbf{Reference} & \textbf{Procurement cost structure} & \textbf{Capacity and prices} & \textbf{Complexity} \\
\midrule
Wagelmans et al.~\cite{wagelmans1992economic}, Aggarwal \& Park~\cite{aggarwal1993improved} & Fixed setup plus linear & Uncapacitated, arbitrary prices & $O(n\log n)$ \\
Federgruen \& Lee~\cite{federgruen1990dynamic} & Setup plus 1-breakpoint all-units discount & Uncapacitated, non-increasing prices & $O(n^3)$ \\
Koca et al.~\cite{koca2014lot} & Piecewise concave ($m$ stationary breakpoints) & Uncapacitated, arbitrary prices & $O(n^{2m+3})$ \\
Ou~\cite{ou2017improved} & Piecewise linear ($m$ stationary breakpoints) & Uncapacitated, arbitrary prices & $O(n^{m+2}\log n)$ \\
Malekian et al.~\cite{malekian2021polynomial} & Setup plus 1-breakpoint all-units discount & Constant capacity, non-increasing prices & $O(n^4)$ \\
Papadopoulos~\cite{papadopoulos2026piecewise} & Piecewise linear ($m$ stationary breakpoints) & Uncapacitated, arbitrary prices & $O(n^{m+2})$ \\
Papadopoulos~\cite{papadopoulos2026monotone} & Linear with 1-breakpoint all-units discount & Time-varying capacity, non-increasing prices & $O(n\log n)$ \\
\textbf{This paper} & \textbf{Linear with 1-breakpoint all-units discount} & \textbf{Uncapacitated, arbitrary prices} & \textbf{$O(n\log n)$} \\
\bottomrule
\end{tabular}
\caption{Comparison of exact time complexities for related single-item lot-sizing models over $n$ periods.}
\label{tab:literature}
\end{table}

For balanced search trees and split/join primitives, we rely on standard augmented tree techniques~\cite{brass2008advanced} and the join-based framework of Blelloch, Ferizovic, and Sun~\cite{blelloch2016join}. All problem-specific invariants, cached thresholds, and lazy propagation operators are self-contained below.

\section{Model, Effective-Price Reduction, and Auxiliary Recurrence}
\label{sec:model}

\subsection{Problem formulation}

Consider a finite horizon of $n \ge 1$ periods indexed by $i = 1, \dots, n$. Let $\mathcal{G} = \mathbb{R}_{\ge 0}$ in the continuous model and $\mathcal{G} = \mathbb{Z}_{\ge 0}$ in the integral model. In period $i$, deterministic demand is $d_i \in \mathcal{G}$, the holding cost per unit of inventory remaining after demand is $h_i \ge 0$, and the common all-units discount threshold is $Q \in \mathcal{G}$ with $Q > 0$. The regular unit price $p_{1,i}$ and discounted unit price $p_{2,i}$ are arbitrary finite real numbers satisfying only the within-period discount ordering
\begin{equation}
  0 \le p_{2,i} \le p_{1,i} < +\infty, \qquad i = 1, \dots, n.
  \label{eq:price-tier}
\end{equation}
No temporal monotonicity is assumed across periods. The procurement cost of ordering $q \in \mathcal{G}$ units in period $i$ is
\begin{equation}
  P_i(q) =
  \begin{cases}
    p_{1,i} q, & 0 \le q < Q, \\
    p_{2,i} q, & q \ge Q.
  \end{cases}
  \label{eq:procurement-cost}
\end{equation}
Note that $P_i(0) = 0$ and $P_i$ is subadditive: $P_i(q + r) \le P_i(q) + P_i(r)$ for all $q, r \ge 0$, because $q + r < Q$ keeps both summands in the regular tier whereas $q + r \ge Q$ prices the combined batch at $p_{2,i} \le p_{1,i}$.

For any prefix horizon $k \in \{1, \dots, n\}$, an order sequence $q = (q_1, \dots, q_k) \in \mathcal{G}^k$ and its induced inventory sequence $I = (I_0, I_1, \dots, I_k) \in \mathcal{G}^{k+1}$ are \emph{feasible} if
\begin{equation}
  I_0 = 0, \qquad I_i = I_{i-1} + q_i - d_i \ge 0, \qquad q_i \ge 0, \qquad i = 1, \dots, k.
  \label{eq:feasibility}
\end{equation}
There is no upper capacity on $I_i$ or $q_i$, and no disposal or resale operation. The total cost on prefix $1, \dots, k$ is
\begin{equation}
  J_k(q, I) = \sum_{i=1}^k \bigl(P_i(q_i) + h_i I_i\bigr).
  \label{eq:objective}
\end{equation}
We denote by $\operatorname{OPT}_k$ the minimum of $J_k(q, I)$ over all feasible schedules on $1, \dots, k$ with free terminal inventory $I_k \ge 0$, writing $\operatorname{OPT} = \operatorname{OPT}_n$ for the full horizon, and by $\operatorname{OPT}_{k,0}$ the minimum of $J_k(q, I)$ subject to the additional terminal constraint $I_k = 0$.

Our complexity bounds use a unit-cost exact real-arithmetic model in which elementary arithmetic operations ($+, -, \times, \div$) and comparisons take $O(1)$ time, supplemented in the integer model $\mathcal{G} = \mathbb{Z}_{\ge 0}$ by $O(1)$-time exact floor and ceiling operations. Throughout, we write $\log n$ for $\max\{1, \log_2 n\}$.

\begin{lemma}[Existence of optimal physical schedules]
\label{lem:physical-attainment}
For every $k \in \{1, \dots, n\}$, both $\operatorname{OPT}_k$ and $\operatorname{OPT}_{k,0}$ are finite and attained by feasible schedules satisfying $\sum_{i=1}^k q_i \le D_k + Q$, where $D_k = \sum_{i=1}^k d_i$.
\end{lemma}

\begin{proof}
Ordering $q_i = d_i$ for each $i = 1, \dots, k$ yields $I_i = 0$ throughout, so both feasible sets are nonempty and have finite infima. For $I_k = 0$, every feasible schedule satisfies $\sum_{i=1}^k q_i = D_k$, confining $(q_1, \dots, q_k)$ to $[0, D_k]^k$. For free $I_k \ge 0$, suppose a feasible schedule has $I_k \ge Q$. Let $j \le k$ be the first period with $\sum_{t=1}^j q_t > D_k$, and put $\ell = D_k - \sum_{t=1}^{j-1} q_t \in [0, q_j)$. Keeping $q_1, \dots, q_{j-1}$, zeroing $q_{j+1}, \dots, q_k$, and replacing $q_j$ by $\ell$ (if $q_j < Q$ or $\ell = 0$) or $\max\{\ell, Q\}$ (if $q_j \ge Q$ and $\ell > 0$) preserves feasibility in $\mathcal{G}^k$, keeps $q'_j$ in its original price tier (or zero), lowers all inventories $I'_t \le I_t$, and achieves $I'_k < Q$ with $J_k(q', I') \le J_k(q, I)$ because $p_{2,t} \ge 0$ and $h_t \ge 0$. Hence we may restrict attention to the bounded set $\sum_{i=1}^k q_i \le D_k + Q$. In $\mathbb{Z}_{\ge 0}$ this set is finite; in $\mathbb{R}_{\ge 0}$ it is compact and each $P_i$ is lower semicontinuous at its downward jump $Q$, so the minimum is attained.
\end{proof}

\subsection{Online effective-price envelope and retroactive order aggregation}
\label{sec:effective-reduction}

For $1 \le t \le i \le n + 1$, define the cumulative holding charge from period $t$ to the start of period $i$ by
\begin{equation}
  H(t, i) = \sum_{m=t}^{i-1} h_m,
  \label{eq:cumulative-holding}
\end{equation}
with $H(i, i) = 0$. Carrying one unit purchased in period $t$ through periods $t, \dots, i-1$ adds exactly $H(t, i)$ to its unit procurement cost when it arrives at period $i$. Define the \emph{effective unit prices} $(\alpha_i, \beta_i)$ and their \emph{origin pointers} $(s(i), r(i))$ recursively in $O(1)$ time per period:
\begin{align}
  \alpha_1 &= p_{1,1}, \quad s(1) = 1; \qquad
  \alpha_i = \min\{\alpha_{i-1} + h_{i-1},\; p_{1,i}\}, \quad
  s(i) = \begin{cases} i, & p_{1,i} \le \alpha_{i-1} + h_{i-1}, \\ s(i-1), & \text{otherwise}, \end{cases}
  \label{eq:effective-alpha} \\
  \beta_1 &= p_{2,1}, \quad r(1) = 1; \qquad
  \beta_i = \min\{\beta_{i-1} + h_{i-1},\; p_{2,i}\}, \quad
  r(i) = \begin{cases} i, & p_{2,i} \le \beta_{i-1} + h_{i-1}, \\ r(i-1), & \text{otherwise}, \end{cases}
  \label{eq:effective-beta}
\end{align}
for $i = 2, \dots, n$. Unrolling the recurrences immediately yields
\begin{equation}
  \alpha_i = \min_{1 \le t \le i} \bigl(p_{1,t} + H(t, i)\bigr) = p_{1, s(i)} + H(s(i), i), \qquad
  \beta_i = \min_{1 \le t \le i} \bigl(p_{2,t} + H(t, i)\bigr) = p_{2, r(i)} + H(r(i), i).
  \label{eq:effective-unrolled}
\end{equation}
Because $0 \le p_{2,t} \le p_{1,t}$ and $h_m \ge 0$, the effective prices satisfy the \emph{non-speculative inequalities}
\begin{equation}
  0 \le \beta_i \le \alpha_i < +\infty \quad (1 \le i \le n), \qquad
  \alpha_{i+1} \le \alpha_i + h_i, \quad \beta_{i+1} \le \beta_i + h_i \quad (1 \le i < n).
  \label{eq:nonspeculative}
\end{equation}

Define the \emph{virtual procurement cost} in period $i$ by
\begin{equation}
  \widehat{P}_i(\hat{q}) =
  \begin{cases}
    \alpha_i \hat{q}, & 0 \le \hat{q} < Q, \\
    \beta_i \hat{q}, & \hat{q} \ge Q,
  \end{cases}
  \label{eq:virtual-procurement}
\end{equation}
and for any feasible virtual schedule $(\hat{q}, \hat{I})$ satisfying \eqref{eq:feasibility} on $1, \dots, k$, define its virtual objective by
\begin{equation}
  \widehat{J}_k(\hat{q}, \hat{I}) = \sum_{i=1}^k \bigl(\widehat{P}_i(\hat{q}_i) + h_i \hat{I}_i\bigr).
  \label{eq:virtual-objective}
\end{equation}
Denote its free-terminal and zero-terminal minima on $1, \dots, k$ by $\widehat{\operatorname{OPT}}_k$ and $\widehat{\operatorname{OPT}}_{k,0}$.

\begin{theorem}[Equivalence of the virtual model and $O(k)$ physical schedule mapping]
\label{thm:effective-equivalence}
For every prefix $k \in \{1, \dots, n\}$,
\begin{equation}
  \operatorname{OPT}_k = \widehat{\operatorname{OPT}}_k \qquad \text{and} \qquad \operatorname{OPT}_{k,0} = \widehat{\operatorname{OPT}}_{k,0}.
  \label{eq:virtual-equivalence}
\end{equation}
Specifically, given any feasible virtual schedule $(\hat{q}_1, \dots, \hat{q}_k) \in \mathcal{G}^k$ with virtual inventories $(\hat{I}_0, \dots, \hat{I}_k)$, assign each positive virtual order $\hat{q}_i > 0$ to the origin period
\begin{equation}
  \tau(i) =
  \begin{cases}
    s(i), & 0 < \hat{q}_i < Q, \\
    r(i), & \hat{q}_i \ge Q,
  \end{cases}
  \label{eq:origin-map}
\end{equation}
and define the physical order sequence $(q_1, \dots, q_k) \in \mathcal{G}^k$ by
\begin{equation}
  q_t = \sum_{\substack{1 \le i \le k \\ \hat{q}_i > 0,\; \tau(i) = t}} \hat{q}_i, \qquad t = 1, \dots, k.
  \label{eq:physical-aggregation}
\end{equation}
Then $(q_1, \dots, q_k)$ is feasible on $1, \dots, k$, induces physical inventories satisfying $I_m \ge \hat{I}_m \ge 0$ for all $1 \le m \le k$ with terminal equality $I_k = \hat{I}_k$, is computable in $O(k)$ time, and satisfies
\begin{equation}
  J_k(q, I) \le \widehat{J}_k(\hat{q}, \hat{I}).
  \label{eq:cost-domination}
\end{equation}
\end{theorem}

\begin{proof}
Since $\alpha_i \le p_{1,i}$ and $\beta_i \le p_{2,i}$ for every $i$, we have $\widehat{P}_i(q_i) \le P_i(q_i)$ for every feasible physical schedule $(q, I)$, giving $\widehat{J}_k(q, I) \le J_k(q, I)$ and hence $\widehat{\operatorname{OPT}}_k \le \operatorname{OPT}_k$ and $\widehat{\operatorname{OPT}}_{k,0} \le \operatorname{OPT}_{k,0}$.

Conversely, fix a feasible virtual schedule $(\hat{q}, \hat{I})$ on $1, \dots, k$ and construct $(q_1, \dots, q_k)$ by \eqref{eq:origin-map}--\eqref{eq:physical-aggregation}. Clearly $q_t \in \mathcal{G}$ because $\mathcal{G}$ is closed under nonnegative addition, and bucket-summing by $\tau(i) \in \{1, \dots, i\}$ takes $O(k)$ time. Because $\tau(i) \le i$ for every positive virtual order, for each $m \in \{1, \dots, k\}$ we have
\[
  \sum_{t=1}^m q_t = \sum_{\substack{1 \le i \le k \\ \hat{q}_i > 0,\; \tau(i) \le m}} \hat{q}_i
  = \sum_{i=1}^m \hat{q}_i + \sum_{\substack{m < i \le k \\ \hat{q}_i > 0,\; \tau(i) \le m}} \hat{q}_i.
\]
Subtracting $D_m = \sum_{t=1}^m d_t$ yields the physical inventory at the end of period $m$:
\begin{equation}
  I_m = \hat{I}_m + \sum_{\substack{m < i \le k \\ \hat{q}_i > 0,\; \tau(i) \le m}} \hat{q}_i \ge \hat{I}_m \ge 0.
  \label{eq:inventory-lift}
\end{equation}
Thus $(q, I)$ satisfies \eqref{eq:feasibility}, and for $m = k$ the second sum in \eqref{eq:inventory-lift} is empty, proving $I_k = \hat{I}_k$.

Multiplying \eqref{eq:inventory-lift} by $h_m$, summing over $m = 1, \dots, k$, and exchanging the order of summation gives
\begin{equation}
  \sum_{m=1}^k h_m I_m
  = \sum_{m=1}^k h_m \hat{I}_m + \sum_{\substack{1 \le i \le k \\ \hat{q}_i > 0}} \hat{q}_i \sum_{m=\tau(i)}^{i-1} h_m
  = \sum_{m=1}^k h_m \hat{I}_m + \sum_{\substack{1 \le i \le k \\ \hat{q}_i > 0}} H(\tau(i), i)\,\hat{q}_i.
  \label{eq:holding-decomposition}
\end{equation}
Next, fix any period $t \in \{1, \dots, k\}$ and let $T_t = \{i \in \{1, \dots, k\} : \hat{q}_i > 0,\; \tau(i) = t\}$, so $q_t = \sum_{i \in T_t} \hat{q}_i$.
\begin{itemize}
\item If $T_t = \emptyset$, then $q_t = 0$ and $P_t(q_t) = 0$.
\item If $0 < q_t < Q$, then every $i \in T_t$ satisfies $0 < \hat{q}_i < Q$, so $t = \tau(i) = s(i)$ and \eqref{eq:effective-unrolled} gives $p_{1,t} = \alpha_i - H(t, i)$. Hence
\[
  P_t(q_t) = p_{1,t} q_t = \sum_{i \in T_t} \bigl(\alpha_i - H(\tau(i), i)\bigr)\hat{q}_i = \sum_{i \in T_t} \bigl(\widehat{P}_i(\hat{q}_i) - H(\tau(i), i)\hat{q}_i\bigr).
\]
\item If $q_t \ge Q$, the aggregated physical order qualifies for the discounted tier $p_{2,t}$, so $P_t(q_t) = p_{2,t} q_t$. For each $i \in T_t$, either $\hat{q}_i \ge Q$, in which case $t = r(i)$ and $p_{2,t}\hat{q}_i = (\beta_i - H(t, i))\hat{q}_i = \widehat{P}_i(\hat{q}_i) - H(\tau(i), i)\hat{q}_i$; or $0 < \hat{q}_i < Q$, in which case $t = s(i)$ and $p_{2,t} \le p_{1,t} = \alpha_i - H(t, i)$, so $p_{2,t}\hat{q}_i \le \widehat{P}_i(\hat{q}_i) - H(\tau(i), i)\hat{q}_i$. Summing over $i \in T_t$ gives
\[
  P_t(q_t) \le \sum_{i \in T_t} \bigl(\widehat{P}_i(\hat{q}_i) - H(\tau(i), i)\hat{q}_i\bigr).
\]
\end{itemize}
Summing this bound over $t = 1, \dots, k$ and adding \eqref{eq:holding-decomposition} cancels every $H(\tau(i), i)\hat{q}_i$ term, yielding \eqref{eq:cost-domination} and completing the proof of \eqref{eq:virtual-equivalence}.
\end{proof}

\begin{remark}[Offline zero-holding-cost reduction via summation by parts]
\label{rem:offline-reduction}
When the entire horizon $n$ is known in advance, substituting $I_i = \sum_{t=1}^i q_t - D_i$ into \eqref{eq:objective} and setting $H_t = \sum_{i=t}^n h_i$ yields $J_n(q, I) = \sum_{t=1}^n \widetilde{P}_t(q_t) - \sum_{t=1}^n H_t d_t$, where $\widetilde{P}_t$ has zero holding costs and unit prices $\tilde{a}_t = p_{1,t} + H_t$, $\tilde{b}_t = p_{2,t} + H_t$. Taking prefix minima $\hat{a}_t = \min_{1 \le s \le t} \tilde{a}_s$ and $\hat{b}_t = \min_{1 \le r \le t} \tilde{b}_r$ reduces the full-horizon free-terminal problem and all zero-terminal prefix subproblems directly to an offline zero-holding-cost instance with non-increasing prices ($\hat{a}_{t+1} \le \hat{a}_t$, $\hat{b}_{t+1} \le \hat{b}_t$). Working directly with the online effective prices $(\alpha_i, \beta_i, h_i)$ in \eqref{eq:effective-alpha}--\eqref{eq:effective-beta} avoids reading future holding rates $h_{i+1}, \dots, h_n$ and preserves free-terminal prefix values $\operatorname{OPT}_k$ at every intermediate period $k$ as well.
\end{remark}

\subsection{A  normal form under non-speculative prices}
\label{sec:normal-form}

We now establish a simultaneous structural normal form for optimal schedules of the virtual model $(\alpha_i, \beta_i, h_i)$ using only the non-speculative inequalities \eqref{eq:nonspeculative}.

\begin{lemma}[Postponement under non-speculative prices]
\label{lem:postponement}
For all $q, r \ge 0$ and $1 \le i < n$, the virtual procurement functions \eqref{eq:virtual-procurement} satisfy
\begin{equation}
  \widehat{P}_{i+1}(q) \le \widehat{P}_i(q) + h_i q \qquad \text{and} \qquad
  \widehat{P}_{i+1}(q + r) \le \widehat{P}_i(q) + h_i q + \widehat{P}_{i+1}(r).
  \label{eq:nonspeculative-postponement}
\end{equation}
\end{lemma}

\begin{proof}
If $0 \le q < Q$, then $\widehat{P}_{i+1}(q) = \alpha_{i+1} q \le (\alpha_i + h_i)q = \widehat{P}_i(q) + h_i q$ by \eqref{eq:nonspeculative}. If $q \ge Q$, then $\widehat{P}_{i+1}(q) = \beta_{i+1} q \le (\beta_i + h_i)q = \widehat{P}_i(q) + h_i q$. Combining this inequality with subadditivity $\widehat{P}_{i+1}(q + r) \le \widehat{P}_{i+1}(q) + \widehat{P}_{i+1}(r)$ gives the second inequality in \eqref{eq:nonspeculative-postponement}.
\end{proof}

\begin{lemma}[Tier-preserving trimming with holding-cost credit]
\label{lem:trimming}
Let $(\hat{q}, \hat{I})$ be a feasible virtual schedule through period $m \ge 1$ with terminal inventory $w = \hat{I}_m > 0$, and let
\begin{equation}
  0 \le \gamma \le \min_{1 \le t \le m} \bigl(\beta_t + H(t, m + 1)\bigr).
  \label{eq:gamma-bound}
\end{equation}
There exists a feasible virtual schedule $(\hat{q}', \hat{I}')$ through period $m$ ending with inventory $s = \hat{I}'_m < Q$ such that $\hat{I}'_r \le \hat{I}_r$ for every $1 \le r \le m$ and
\begin{equation}
  \widehat{J}_m(\hat{q}', \hat{I}') \le \widehat{J}_m(\hat{q}, \hat{I}) - \gamma(w - s).
  \label{eq:trimming-saving}
\end{equation}
In particular, if $m = i - 1 < n$, \eqref{eq:nonspeculative} implies $\beta_t + H(t, i) \ge \beta_i$ for all $1 \le t \le i - 1$, so \eqref{eq:trimming-saving} holds with $\gamma = \beta_i$.
\end{lemma}

\begin{proof}
Write $D_r = \sum_{t=1}^r d_t$ and $\hat{S}_r = \sum_{t=1}^r \hat{q}_t$, with $D_0 = \hat{S}_0 = 0$. Since $\hat{S}_m = D_m + w > D_m$, let $j \in \{1, \dots, m\}$ be the first index with $\hat{S}_j > D_m$, and set $\ell = D_m - \hat{S}_{j-1} \in [0, \hat{q}_j)$. Define $(\hat{q}'_1, \dots, \hat{q}'_m)$ by keeping $\hat{q}'_t = \hat{q}_t$ for $t < j$, setting $\hat{q}'_t = 0$ for $t > j$, and putting
\begin{equation}
  \hat{q}'_j =
  \begin{cases}
    \ell, & \hat{q}_j < Q \text{ or } \ell = 0, \\
    \max\{\ell, Q\}, & \hat{q}_j \ge Q \text{ and } \ell > 0.
  \end{cases}
  \label{eq:trimmed-order}
\end{equation}
Then $\hat{q}' \in \mathcal{G}^m$, $\ell \le \hat{q}'_j \le \hat{q}_j$, and $s = \hat{q}'_j - \ell < Q$. For $r < j$ the schedule is unchanged; for $j \le r \le m$, $\hat{S}'_r = \hat{S}_{j-1} + \hat{q}'_j = D_m + s \ge D_r$ and $\hat{S}'_r \le \hat{S}_r$, so $0 \le \hat{I}'_r \le \hat{I}_r$.

Let $\delta_t = \hat{q}_t - \hat{q}'_t \ge 0$ for $1 \le t \le m$ (so $\delta_t = 0$ for $t < j$ and $\sum_{t=j}^m \delta_t = w - s$). Because \eqref{eq:trimmed-order} keeps any positive $\hat{q}'_j$ in the same price tier as $\hat{q}_j$, every period $t \in \{j, \dots, m\}$ satisfies $\widehat{P}_t(\hat{q}_t) - \widehat{P}_t(\hat{q}'_t) \ge \beta_t \delta_t$. Furthermore, for every $r \in \{1, \dots, m\}$, $\hat{I}_r - \hat{I}'_r = \hat{S}_r - \hat{S}'_r = \sum_{t=1}^r \delta_t$, so the holding-cost reduction equals
\[
  \sum_{r=1}^m h_r(\hat{I}_r - \hat{I}'_r) = \sum_{r=1}^m h_r \sum_{t=1}^r \delta_t = \sum_{t=j}^m \delta_t \sum_{r=t}^m h_r = \sum_{t=j}^m \delta_t H(t, m + 1).
\]
Adding procurement and holding savings gives
\[
  \widehat{J}_m(\hat{q}, \hat{I}) - \widehat{J}_m(\hat{q}', \hat{I}') \ge \sum_{t=j}^m \delta_t \bigl(\beta_t + H(t, m + 1)\bigr) \ge \gamma \sum_{t=j}^m \delta_t = \gamma(w - s).
\]
Finally, iterating $\beta_r + h_r \ge \beta_{r+1}$ across $r = t, \dots, i - 1$ gives $\beta_t + H(t, i) \ge \beta_i$, proving admissibility of $\gamma = \beta_i$.
\end{proof}

\begin{theorem}[Simultaneous normal form for the virtual problem]
\label{thm:normal-form}
For every horizon $m \in \{1, \dots, n\}$, the free-terminal virtual problem admits an optimal feasible schedule $(\hat{q}_1, \dots, \hat{q}_m)$ satisfying simultaneously
\begin{align}
  \hat{q}_i > 0 &\implies \hat{I}_{i-1} < d_i, && i = 1, \dots, m, \label{eq:normal-purchase} \\
  \hat{I}_{i-1} &< 2Q, && i = 1, \dots, m, \label{eq:normal-incoming} \\
  \hat{I}_m &< Q. && \label{eq:normal-terminal}
\end{align}
For the zero-terminal virtual problem ($\hat{I}_m = 0$), there exists an optimal feasible schedule satisfying \eqref{eq:normal-purchase}--\eqref{eq:normal-incoming} simultaneously on $1, \dots, m$.
\end{theorem}

\begin{proof}
By Lemma~\ref{lem:physical-attainment} applied to $(\alpha_i, \beta_i, h_i)$, the set of optimal virtual schedules on $1, \dots, m$ (restricted to $\sum_{i=1}^m \hat{q}_i \le D_m + Q$ in the free-terminal case, since Lemma~\ref{lem:trimming} with $\gamma = 0$ trims any optimal schedule with $\hat{I}_m \ge Q$ without raising cost) is nonempty and compact in $\mathbb{R}_{\ge 0}^m$ (or finite in $\mathbb{Z}_{\ge 0}^m$). Among all optimal virtual schedules, choose one minimizing the linear inventory potential
\begin{equation}
  \Phi(\hat{I}) = \sum_{t=1}^m \hat{I}_t.
  \label{eq:potential}
\end{equation}

First, suppose $i < m$, $\hat{I}_{i-1} = u \ge d_i$, and $\hat{q}_i = q > 0$. Set $r = \hat{q}_{i+1}$ and replace $(\hat{q}_i, \hat{q}_{i+1})$ by $(0, q + r)$. Then $\hat{I}'_i = u - d_i = \hat{I}_i - q \ge 0$, while $\hat{I}'_i + (q + r) = \hat{I}_i + r$ leaves $\hat{I}_{i+1}, \dots, \hat{I}_m$ unchanged. By Lemma~\ref{lem:postponement}, total procurement plus period-$i$ holding cost changes by $\widehat{P}_{i+1}(q + r) - \widehat{P}_i(q) - \widehat{P}_{i+1}(r) - h_i q \le 0$, whereas $\Phi(\hat{I})$ strictly decreases by $q > 0$, contradicting minimality of $\Phi$. At $i = m$, $\hat{I}_{m-1} \ge d_m$ and $\hat{q}_m > 0$ is impossible when $\hat{I}_m = 0$, and in the free-terminal case setting $\hat{q}'_m = 0$ preserves $\hat{I}'_m \ge 0$, lowers cost weakly, and strictly decreases $\Phi$. Thus \eqref{eq:normal-purchase} holds.

Second, suppose $w = \hat{I}_{i-1} \ge 2Q$ for some $2 \le i \le m$. Apply Lemma~\ref{lem:trimming} through period $i - 1$ with $\gamma = \beta_i$ to obtain $(\hat{q}'_1, \dots, \hat{q}'_{i-1})$ ending at $s = \hat{I}'_{i-1} < Q$ with $\widehat{J}_{i-1}(\hat{q}', \hat{I}') \le \widehat{J}_{i-1}(\hat{q}, \hat{I}) - \beta_i(w - s)$. Set $\hat{q}'_i = \hat{q}_i + w - s$ and keep $\hat{q}'_t = \hat{q}_t$ for $t > i$. Because $w \ge 2Q$ and $s < Q$, we have $\hat{q}'_i > Q$, so $\widehat{P}_i(\hat{q}'_i) - \widehat{P}_i(\hat{q}_i) = \beta_i(\hat{q}_i + w - s) - \widehat{P}_i(\hat{q}_i) \le \beta_i(w - s)$. Since $s + \hat{q}'_i = w + \hat{q}_i$, inventories from period $i$ onward are identical, earlier inventories weakly decrease, and $\hat{I}'_{i-1} = s < w$ strictly decreases $\Phi(\hat{I})$ without increasing total cost---a contradiction. Since $\hat{I}_0 = 0$, \eqref{eq:normal-incoming} holds for all $i$.

Third, in the free-terminal case, $\hat{I}_m \ge Q$ is ruled out by applying Lemma~\ref{lem:trimming} on $1, \dots, m$ with $\gamma = 0$, which yields $\hat{I}'_m < Q$ and strictly smaller $\Phi$. Because every contradiction is derived against the same minimizer of $\Phi$, \eqref{eq:normal-purchase}--\eqref{eq:normal-terminal} hold simultaneously.
\end{proof}

\subsection{The auxiliary dynamic program and single-crossover property}
\label{sec:auxiliary-dp}

Define the bounded state and departure domains
\begin{equation}
  K_1 = 0, \qquad K_{i+1} = 2Q, \qquad Y_i = d_i + 2Q, \qquad i = 1, \dots, n.
  \label{eq:aux-domains}
\end{equation}
The incoming auxiliary value function $A_i$ is supported on $[0, K_i] \cap \mathcal{G}$ (extended by $+\infty$ elsewhere), initialized by $A_1(0) = 0$. For $i = 1, \dots, n$ and $y \in [0, Y_i] \cap \mathcal{G}$, define
\begin{align}
  O_i(y) &= \inf_{0 \le u \le y} \bigl\{A_i(u) + \alpha_i(y - u)\bigr\}, \label{eq:regular-op} \\
  N_i(y) &= \inf_{\substack{0 \le u < d_i \\ u + Q \le y}} \bigl\{A_i(u) + \beta_i(y - u)\bigr\}, \label{eq:discount-op} \\
  F_i(y) &= \min\{O_i(y), N_i(y)\}, \label{eq:departure-op} \\
  A_{i+1}(z) &= F_i(d_i + z) + h_i z, \qquad z \in [0, K_{i+1}] \cap \mathcal{G}, \label{eq:aux-update}
\end{align}
where $\inf \emptyset = +\infty$ and predecessor variables $u$ range over $\mathcal{G}$. In this auxiliary model, a regular action may purchase any quantity $y - u \ge 0$ at unit rate $\alpha_i$ (even if $y - u \ge Q$), whereas a discounted action pays $\beta_i$ per unit, requires $y - u \ge Q$, and is permitted only when $u < d_i$.

\begin{theorem}[Exactness of the auxiliary dynamic program]
\label{thm:aux-exactness}
For every prefix $k \in \{1, \dots, n\}$, the auxiliary recurrence \eqref{eq:regular-op}--\eqref{eq:aux-update} satisfies
\begin{equation}
  A_{k+1}(0) = \operatorname{OPT}_{k,0} \qquad \text{and} \qquad
  \min_{0 \le z \le K_{k+1}} A_{k+1}(z) = \operatorname{OPT}_k,
  \label{eq:aux-equivalence}
\end{equation}
with the second minimum attained at some $z < Q$. In particular, $\min_{0 \le z \le 2Q} A_{n+1}(z) = \operatorname{OPT}$.
\end{theorem}

\begin{proof}
By Theorem~\ref{thm:effective-equivalence}, it suffices to compare $A_{k+1}$ with $\widehat{\operatorname{OPT}}_{k,0}$ and $\widehat{\operatorname{OPT}}_k$. Every auxiliary trajectory through period $k$ defines a feasible virtual schedule $(\hat{q}_1, \dots, \hat{q}_k)$ whose auxiliary procurement charges satisfy $\alpha_i \hat{q}_i \ge \widehat{P}_i(\hat{q}_i)$ on regular steps and $\beta_i \hat{q}_i = \widehat{P}_i(\hat{q}_i)$ on discounted steps (where $\hat{q}_i \ge Q$). Thus $A_{k+1}(0) \ge \widehat{\operatorname{OPT}}_{k,0}$ and $\inf_{0 \le z \le 2Q} A_{k+1}(z) \ge \widehat{\operatorname{OPT}}_k$.

Conversely, let $(\hat{q}_1, \dots, \hat{q}_k)$ be the optimal virtual schedule given by Theorem~\ref{thm:normal-form} for prefix $k$ under either terminal convention. Every step with $0 \le \hat{q}_i < Q$ uses the regular operator \eqref{eq:regular-op} at exact cost $\widehat{P}_i(\hat{q}_i) = \alpha_i \hat{q}_i$; every step with $\hat{q}_i \ge Q$ has $\hat{I}_{i-1} < d_i$ by \eqref{eq:normal-purchase} and uses the discounted operator \eqref{eq:discount-op} at exact cost $\widehat{P}_i(\hat{q}_i) = \beta_i \hat{q}_i$. Moreover, $\hat{I}_{i-1} < 2Q$ for all $1 \le i \le k$ and $\hat{I}_k < Q$ (or $\hat{I}_k = 0$), so every visited state lies strictly inside $[0, K_i]$ and every departure level $\hat{I}_{i-1} + \hat{q}_i = d_i + \hat{I}_i$ lies inside $[0, Y_i]$. Hence the auxiliary recurrence evaluates this schedule at exact cost $\widehat{\operatorname{OPT}}_k$ (or $\widehat{\operatorname{OPT}}_{k,0}$), proving equality and attainment below $Q$.
\end{proof}

For an operator notation on a function $C$ extended by $+\infty$ outside its domain, write $(\mathcal{O}_a C)(y) = \inf_{0 \le u \le y}\{C(u) + a(y - u)\}$ and $(\mathcal{D}_{b,d} C)(y) = \inf_{0 \le u < d,\, u + Q \le y}\{C(u) + b(y - u)\}$.

\begin{lemma}[Closure and distributivity]
\label{lem:closure}
For any $a \ge b \ge 0$ and $Y \ge d \ge 0$, restricted to $y \in [0, Y] \cap \mathcal{G}$,
\begin{equation}
  \mathcal{O}_a \mathcal{O}_a C = \mathcal{O}_a C, \qquad
  \mathcal{D}_{b,d} \mathcal{O}_a C = \mathcal{D}_{b,d} C, \qquad
  \mathcal{O}_a \mathcal{D}_{b,d} C = \mathcal{D}_{b,d} C,
  \label{eq:closure-identities}
\end{equation}
and both operators distribute over pointwise infima. Consequently, $F = \min\{\mathcal{O}_a C, \mathcal{D}_{b,d} C\}$ satisfies $\mathcal{O}_a F = F$.
\end{lemma}

\begin{proof}
For $\mathcal{O}_a \mathcal{O}_a C$, $C(v) + a(u - v) + a(y - u) = C(v) + a(y - v)$ for $v \le u \le y$. For $\mathcal{D}_{b,d}\mathcal{O}_a C$, any $v \le u < d$ with $u + Q \le y$ satisfies $v < d$, $v + Q \le y$, and $C(v) + a(u - v) + b(y - u) = C(v) + b(y - v) + (a - b)(u - v) \ge C(v) + b(y - v)$, with equality at $u = v$. For $\mathcal{O}_a \mathcal{D}_{b,d} C$, any $u < d$ with $u + Q \le v \le y$ satisfies $C(u) + b(v - u) + a(y - v) = C(u) + b(y - u) + (a - b)(y - v) \ge C(u) + b(y - u)$, with equality at $v = y$. All intermediate states lie in $[0, Y]$ when $y \le Y$ and $Y \ge d$. Distributivity over pointwise infima gives $\mathcal{O}_a F = F$.
\end{proof}

\begin{proposition}[Piecewise-affine slopes and attainment]
\label{prop:local-slopes}
For each period $i \in \{1, \dots, n\}$, $A_i$, $O_i$, $N_i$, and $F_i$ have finite piecewise-affine descriptions on their finite domains, every affine piece has slope at least $\beta_i$, and every finite infimum in \eqref{eq:regular-op}--\eqref{eq:discount-op} is attained. In the continuous model $\mathcal{G} = \mathbb{R}_{\ge 0}$, the functions are right-continuous on the interior of their finite domains with no upward jumps.
\end{proposition}

\begin{proof}
At $i = 1$, $A_1$ is supported on $\{0\}$, $O_1(y) = \alpha_1 y$ has slope $\alpha_1 \ge \beta_1$, and $N_1(y) = \beta_1 y$ (for $d_1 > 0$, $y \ge Q$) has slope $\beta_1$. Suppose the claim holds for $A_i$. Since every piece of $A_i$ has slope $p \ge \beta_i$, $\mathcal{O}_{\alpha_i} A_i$ produces pieces of slope $\min\{p, \alpha_i\} \ge \beta_i$. Within each piece of $A_i$, $u \mapsto A_i(u) - \beta_i u$ is nondecreasing, so its infimum over $0 \le u < d_i$, $u \le y - Q$ is attained at one of finitely many left endpoints of pieces of $A_i$ strictly below $d_i$ (never requiring an unattained open right endpoint $d_i$). Thus $N_i$ is piecewise affine of slope $\beta_i$ with finitely many downward jumps, and $F_i = \min\{O_i, N_i\}$ preserves slopes $\ge \beta_i$, right-continuity, and lower semicontinuity. Finally, $A_{i+1}(z) = F_i(d_i + z) + h_i z$ shifts all slopes by $+h_i$, so every piece of $A_{i+1}$ has slope at least $\beta_i + h_i \ge \beta_{i+1}$ by \eqref{eq:nonspeculative}.
\end{proof}

\begin{proposition}[Normalized high-inventory invariant]
\label{prop:high-invariant}
For every period $i \in \{1, \dots, n\}$:
\begin{enumerate}
\item $u \mapsto A_i(u) - \beta_i u$ is nondecreasing on $[Q, K_i] \cap \mathcal{G}$;
\item $y \mapsto O_i(y) - \beta_i y$ is nondecreasing on $[Q, Y_i] \cap \mathcal{G}$;
\item $y \mapsto N_i(y) - \beta_i y$ is nonincreasing on $[Q, Y_i] \cap \mathcal{G}$ (with $+\infty$ allowed prior to activation).
\end{enumerate}
\end{proposition}

\begin{proof}
Claim 3 holds directly from $N_i(y) - \beta_i y = \inf_{0 \le u < d_i,\, u \le y - Q}\{A_i(u) - \beta_i u\}$, whose feasible set expands with $y$. Next, suppose Claim 1 holds at period $i$ and fix $Q \le x < y \le Y_i$. For any finite predecessor $u \le y$ in $O_i(y)$: if $u \le x$, then $A_i(u) + \alpha_i(y - u) - \beta_i y = A_i(u) + \alpha_i(x - u) - \beta_i x + (\alpha_i - \beta_i)(y - x) \ge O_i(x) - \beta_i x$; if $u > x \ge Q$, Claim 1 gives $A_i(u) - \beta_i u \ge A_i(x) - \beta_i x \ge O_i(x) - \beta_i x$, and adding $(\alpha_i - \beta_i)(y - u) \ge 0$ yields the same lower bound. Taking the infimum over $u$ proves Claim 2.

We prove Claim 1 by induction on $i$. At $i = 1$, $[Q, K_1] = [Q, 0] = \emptyset$, so Claim 1 is vacuous. Assuming Claim 1 at period $i$, Claim 2 holds at period $i$. For every $z \in [Q, K_{i+1}] \cap \mathcal{G}$, the target $y = d_i + z$ satisfies $y - Q = d_i + (z - Q) \ge d_i$, which activates every predecessor $u < d_i$ simultaneously. Writing $M_i = \inf_{0 \le u < d_i}\{A_i(u) - \beta_i u\} \in (-\infty, +\infty]$, we obtain for $i < n$:
\begin{equation}
  A_{i+1}(z) - \beta_{i+1} z
  = \min\bigl\{O_i(d_i + z) - \beta_i(d_i + z),\; M_i\bigr\} + \beta_i d_i + (\beta_i + h_i - \beta_{i+1})z.
  \label{eq:high-induction}
\end{equation}
By Claim 2, $z \mapsto O_i(d_i + z) - \beta_i(d_i + z)$ is nondecreasing for $z \ge Q$; taking its pointwise minimum with the constant $M_i$ preserves monotonicity; and $(\beta_i + h_i - \beta_{i+1})z$ is nondecreasing by the non-speculative inequality $\beta_{i+1} \le \beta_i + h_i$ in \eqref{eq:nonspeculative}. This completes the induction.
\end{proof}

\begin{theorem}[Single-crossover property]
\label{thm:single-crossing}
For every period $i$ and all $Q \le x \le y \le Y_i$ in $\mathcal{G}$,
\begin{equation}
  N_i(x) \le O_i(x) \implies N_i(y) \le O_i(y),
  \label{eq:single-crossing}
\end{equation}
and if $N_i(x) < O_i(x)$, then $N_i(y) < O_i(y)$. Assigning ties to $N_i$ makes the winning region of $N_i$ on $[\max\{Q, d_i\}, Y_i]$ a contiguous suffix.
\end{theorem}

\begin{proof}
Notice that $O_i$ is finite throughout $[0, Y_i]$ because $A_i(0)$ is finite. If $N_i(x) \le O_i(x)$, then $N_i(x) < +\infty$, and Proposition~\ref{prop:high-invariant} gives
\[
  N_i(y) - \beta_i y \le N_i(x) - \beta_i x \le O_i(x) - \beta_i x \le O_i(y) - \beta_i y,
\]
with strictness preserved when the middle inequality is strict.
\end{proof}

\section{Normalized Candidates and Certified Frontier Operations}
\label{sec:representation}

To maintain $A_i$, $O_i$, $N_i$, and $F_i$ without enumerating quantities up to $Q$ or $d_i$, we represent each frontier as the lower envelope of ordered one-kink candidate functions. Throughout, write $x_+ = \max\{x, 0\}$.

\subsection{Normal-form candidate families}

\begin{definition}[Normalized candidate family]
\label{def:normal-form}
Fix a common regular comparator price $c \ge 0$ and a domain bound $U \in \mathcal{G}$. A candidate $j$ is a tuple $(f_j, v_j, \pi_j, t_j)$ with anchor $f_j \in \mathcal{G}$, finite anchor value $v_j \in \mathbb{R}$, legacy slope label $\pi_j \in [0, +\infty]$ (with $\min\{+\infty, c\} = c$), and legacy endpoint $t_j \in \mathcal{G}$, defining the curve
\begin{equation}
  E_j(x; c) =
  \begin{cases}
    +\infty, & x < f_j, \\[3pt]
    v_j + \min\{\pi_j, c\}\min\{x - f_j, t_j - f_j\} + c(x - t_j)_+, & x \ge f_j.
  \end{cases}
  \label{eq:candidate-curve}
\end{equation}
A nonempty family of $m$ candidates represents $C(x) = \min_{1 \le j \le m} E_j(x; c)$ on $[0, U] \cap \mathcal{G}$ (and $+\infty$ outside; the empty family represents $+\infty$ everywhere) and is in \emph{normal form} if
\begin{equation}
  0 \le f_j \le t_j \le f_{j+1} \quad (1 \le j < m), \qquad f_m \le t_m \le U, \qquad \pi_1 \ge \pi_2 \ge \dots \ge \pi_m,
  \label{eq:interleaving}
\end{equation}
coincident pure-linear candidates ($f_j = t_j = f_{j+1} = t_{j+1}$) are coalesced to retain a single minimum-value candidate, and the \emph{pruned prefix-certificate condition} holds:
\begin{equation}
  E_{j-1}(t_j; c) > E_j(t_j; c), \qquad j = 2, \dots, m.
  \label{eq:adjacent-cert}
\end{equation}
\end{definition}

Coalescing identical pure-linear candidates ensures that the lexicographic pairs $(t_j, f_j)$ are strictly increasing with $j$. In the continuous model, if candidate $j+1$ introduces a downward jump at $f_{j+1} = t_j$, evaluation of \eqref{eq:adjacent-cert} at $t_j$ compares the continuous formulas directly, while point queries evaluate both neighbors so the jump at $f_{j+1}$ is respected.

\begin{lemma}[Checkpoint domination and two-neighbor queries]
\label{lem:two-neighbors}
In any family satisfying \eqref{eq:interleaving}:
\begin{enumerate}
\item For $j < k$, $x \mapsto E_j(x; c) - E_k(x; c)$ is nondecreasing on $[f_k, t_k]$ and constant on $[t_k, +\infty)$. Hence $E_j(t_k; c) \le E_k(t_k; c)$ if and only if $E_j(x; c) \le E_k(x; c)$ for all $x \ge f_k$.
\item If \eqref{eq:adjacent-cert} holds, then $E_h(x; c) > E_j(x; c)$ for every $h < j$ and every $x \ge t_j$.
\item Consequently, for any pruned nonempty family and $x \in [0, U] \cap \mathcal{G}$, let $j$ be the largest index with $t_j \le x$. If no such $j$ exists, $C(x) = E_1(x; c)$; if $j = m$, $C(x) = E_m(x; c)$; and if $1 \le j < m$,
\begin{equation}
  C(x) = \min\bigl\{E_j(x; c),\; E_{j+1}(x; c)\bigr\}.
  \label{eq:two-neighbor-eval}
\end{equation}
\end{enumerate}
\end{lemma}

\begin{proof}
For $j < k$ and $x \ge f_k \ge t_j$, $E_j(x; c)$ has slope $c$, whereas $E_k(x; c)$ has slope $\min\{\pi_k, c\} \le c$ on $[f_k, t_k]$ and slope $c$ on $[t_k, +\infty)$ (and even if $f_k = t_j$, no earlier interval exists in $[f_k, +\infty)$). Thus $E_j - E_k$ attains its maximum over $[f_k, +\infty)$ at $t_k$, proving Claim 1. For Claim 2, for every $\ell < j$ and $x \ge t_j \ge t_{\ell+1}$, $E_\ell(x; c) - E_{\ell+1}(x; c) = E_\ell(t_{\ell+1}; c) - E_{\ell+1}(t_{\ell+1}; c) > 0$; summing over $\ell = h, \dots, j-1$ gives $E_h(x; c) > E_j(x; c)$. For Claim 3, Claim 2 eliminates all $h < j$ at $x \ge t_j$, while $t_{j+1} > x$ implies $f_k \ge t_{k-1} \ge t_{j+1} > x$ for all $k \ge j + 2$, making those candidates $+\infty$ at $x$.
\end{proof}

\subsection{Regular closure and exact three-regime comparator thresholds}

\begin{lemma}[Regular closure via comparator lowering]
\label{lem:regular-closure}
Let $C$ be represented on $[0, U]$ at comparator $c$, and let $0 \le a \le c$. For every $y \ge 0$ in $\mathcal{G}$,
\begin{equation}
  \inf_{0 \le u \le \min\{y, U\}} \bigl\{C(u) + a(y - u)\bigr\} = \min_{1 \le j \le m} E_j(y; a).
  \label{eq:comparator-lowering}
\end{equation}
\end{lemma}

\begin{proof}
Distribute the infimum over $j = 1, \dots, m$. Fix $j$ and put $s = \min\{\pi_j, c\}$. On $[f_j, U]$, $u \mapsto E_j(u; c) - a u$ has slope $s - a$ on $[f_j, t_j]$ and slope $c - a \ge 0$ on $[t_j, U]$. For $y \ge f_j$, its minimum over $u \in [f_j, \min\{y, U\}]$ is attained at $u = f_j \in \mathcal{G}$ if $s \ge a$, and at $u = \min\{y, t_j\} \in \mathcal{G}$ if $s < a$ (since $t_j \le U$). Direct substitution gives $E_j(y; a)$ in both cases.
\end{proof}

\begin{lemma}[Exact three-regime comparator threshold]
\label{lem:threshold}
Consider an adjacent pair $L = (f_L, v_L, \pi_L, t_L)$ and $R = (f_R, v_R, \pi_R, t_R)$ satisfying $f_L \le t_L \le f_R \le t_R$ and $\pi_L \ge \pi_R$. Put
\[
  \lambda_L = t_L - f_L \ge 0, \qquad \Delta_t = t_R - t_L \ge 0, \qquad \lambda_R = t_R - f_R \ge 0, \qquad \delta = f_R - f_L \ge 0.
\]
Note that $\lambda_L + \Delta_t - \lambda_R = \delta$. Extending the right-checkpoint difference algebraically to all $c \in \mathbb{R}$ by
\begin{equation}
  D(c) = E_L(t_R; c) - E_R(t_R; c) = v_L - v_R + \lambda_L \min\{\pi_L, c\} + \Delta_t c - \lambda_R \min\{\pi_R, c\},
  \label{eq:checkpoint-diff}
\end{equation}
the function $D : \mathbb{R} \to \mathbb{R}$ is continuous and nondecreasing, and
\begin{equation}
  \theta = \sup\{c \in \mathbb{R} : D(c) \le 0\} \in [-\infty, +\infty]
  \label{eq:theta-def}
\end{equation}
is computable in $O(1)$ time. For every comparator $c \ge 0$, candidate $L$ dominates candidate $R$ on $[f_R, +\infty)$ if and only if $c \le \theta$.
\end{lemma}

\begin{proof}
For finite labels $\pi_R \le \pi_L$, \eqref{eq:checkpoint-diff} splits into three affine pieces:
\begin{equation}
  D(c) =
  \begin{cases}
    v_L - v_R + \delta c, & c \le \pi_R, \\
    v_L - v_R - \lambda_R \pi_R + (t_R - f_L)c, & \pi_R \le c \le \pi_L, \\
    v_L - v_R + \lambda_L \pi_L - \lambda_R \pi_R + \Delta_t c, & c \ge \pi_L,
  \end{cases}
  \label{eq:three-regimes}
\end{equation}
whose slopes $\delta \ge 0$, $t_R - f_L = \delta + \lambda_R \ge 0$, and $\Delta_t \ge 0$ are all nonnegative, and whose endpoint values match at $\pi_R$ and $\pi_L$ (if $\pi_L = +\infty$ or $\pi_R = +\infty$, omit the corresponding upper regime(s)). Thus $\{c \in \mathbb{R} : D(c) \le 0\}$ is an initial interval $(-\infty, \theta]$ (or $\emptyset$ with $\theta = -\infty$, or $\mathbb{R}$ with $\theta = +\infty$). Solving the affine inequality on each regime handles zero slopes ($\delta = 0$ or $\Delta_t = 0$) directly without division by zero. Equivalence with full-domain domination follows from Lemma~\ref{lem:two-neighbors}(1).
\end{proof}

Retaining finite negative values of $\theta$ (rather than clamping negative roots to $-\infty$) ensures exact invariance under holding-cost additions: adding $h_i x$ to all curves maps $(v_j, \pi_j, c) \mapsto (v_j + h_i f_j, \pi_j + h_i, c + h_i)$, which satisfies $D_{\mathrm{new}}(c + h_i) = D_{\mathrm{old}}(c)$ and shifts every finite threshold by $\theta_{\mathrm{new}} = \theta_{\mathrm{old}} + h_i$, while demand translations $x \mapsto x - d_i$ preserve $\lambda_L, \Delta_t, \lambda_R, \delta$ and leave $\theta$ unchanged.

\begin{theorem}[Completeness of adjacent pruning]
\label{thm:pruning}
Starting from any family satisfying \eqref{eq:interleaving} at comparator $c \ge 0$, repeatedly deleting the right candidate of any adjacent pair with $\theta \ge c$ (and coalescing coincident pure-linear pairs) preserves the lower envelope $C$, maintains \eqref{eq:interleaving}, restores \eqref{eq:adjacent-cert} upon termination, and performs exactly one step per deleted candidate.
\end{theorem}

\begin{proof}
By Lemma~\ref{lem:threshold}, $\theta \ge c \iff D(c) \le 0$, so $E_L(x; c) \le E_R(x; c)$ for all $x \ge f_R$, making $R$ redundant. Deleting $R$ preserves \eqref{eq:interleaving} and affects only the adjacency between $L$ and $\mathrm{succ}(R)$. When $\theta < c$ holds for all remaining adjacent pairs, $D(c) > 0$ holds everywhere, establishing \eqref{eq:adjacent-cert}.
\end{proof}

\subsection{Bulk ray generation, domain clipping, and merge truncation}

\begin{lemma}[Bulk rays from eligible anchors and parallel-record normalization]
\label{lem:bulk-normalization}
Suppose $C$ is represented on $[0, U]$ at comparator $c \ge \beta_i$ with $\pi_j \ge \beta_i$ for every candidate $j$. Then for all $y \in [0, Y_i] \cap \mathcal{G}$,
\begin{equation}
  (\mathcal{D}_{\beta_i, d_i} C)(y) = \min_{\substack{j:\; f_j < d_i \\ f_j + Q \le y}} \bigl\{v_j + \beta_i(y - f_j)\bigr\}.
  \label{eq:anchor-bulk}
\end{equation}
For each eligible candidate ($f_j < d_i$ and $F_j = f_j + Q \le Y_i$), set $V_j = v_j + Q\beta_i$ and intercept $\eta_j = V_j - \beta_i F_j = v_j - \beta_i f_j$. Retaining the minimum-value ray at equal anchors and keeping only strict prefix-record minima of $\eta_j$ in increasing order of $F_j$ produces surviving records $1, \dots, k$ whose lower envelope on $[0, Y_i] \cap \mathcal{G}$ is represented in normal form (at any comparator $c \ge \beta_i$) by $(F_r, V_r, \beta_i, T_r)$ with
\begin{equation}
  T_r =
  \begin{cases}
    F_{r+1}, & 1 \le r < k \text{ when } \mathcal{G} = \mathbb{R}_{\ge 0}, \\
    F_{r+1} - 1, & 1 \le r < k \text{ when } \mathcal{G} = \mathbb{Z}_{\ge 0}, \\
    Y_i, & r = k,
  \end{cases}
  \label{eq:bulk-endpoints}
\end{equation}
and exact adjacent thresholds equal to the anchor secants
\begin{equation}
  \theta_r = R_r = \frac{V_{r+1} - V_r}{F_{r+1} - F_r} < \beta_i, \qquad r = 1, \dots, k - 1.
  \label{eq:bulk-secant}
\end{equation}
\end{lemma}

\begin{proof}
On each candidate $j$, $\min\{\pi_j, c\} \ge \beta_i$ and $c \ge \beta_i$, so $u \mapsto E_j(u; c) - \beta_i u$ is nondecreasing for $u \ge f_j$. Whenever candidate $j$ has a feasible source $u \in [f_j, d_i)$ with $u + Q \le y$, $u = f_j$ is also feasible and attains its minimum, proving \eqref{eq:anchor-bulk}. Because all shifted candidates $V_j + \beta_i(y - F_j) = \eta_j + \beta_i y$ are parallel rays of slope $\beta_i$, a ray is optimal somewhere if and only if its intercept $\eta_j$ sets a strict new minimum in anchor order; record $r$ then wins on $[F_r, F_{r+1})$ (or $\{F_r, \dots, F_{r+1} - 1\}$ in $\mathbb{Z}_{\ge 0}$), and raising its continuation beyond $T_r$ to slope $c \ge \beta_i$ leaves the combined lower envelope unchanged because record $r + 1$ is strictly cheaper from $F_{r+1}$ onward.

Finally, for $c \le \beta_i$, substituting $\pi_r = \pi_{r+1} = \beta_i$ into \eqref{eq:checkpoint-diff} gives $\min\{\beta_i, c\} = c$, so both $T_r$ and $T_{r+1}$ cancel identically in both $\mathbb{R}_{\ge 0}$ and $\mathbb{Z}_{\ge 0}$:
\[
  D_r(c) = V_r - V_{r+1} + \bigl((T_r - F_r) + (T_{r+1} - T_r) - (T_{r+1} - F_{r+1})\bigr)c = V_r - V_{r+1} + (F_{r+1} - F_r)c.
\]
At $c = \beta_i$, $D_r(\beta_i) = \eta_r - \eta_{r+1} > 0$. Monotonicity of $D_r$ implies $D_r(c) > 0$ for all $c \ge \beta_i$ and yields the unique root $\theta_r = R_r < \beta_i$ in \eqref{eq:bulk-secant}.
\end{proof}

\begin{lemma}[Domain clipping and merge-boundary truncation]
\label{lem:clipping}
Let a normal-form family represent $C$ on $[0, U]$ at comparator $c$.
\begin{enumerate}
\item \emph{Lower-bound clipping to $[\ell, U]$ ($0 \le \ell \le U$):} Let $j$ be the last candidate with $t_j \le \ell$ (if one exists). Discarding indices below $j$, replacing $j$ by the pure-linear boundary candidate $(\ell, E_j(\ell; c), \pi_j, \ell)$, and---if $j + 1$ (or $1$, when $j$ does not exist) has $f < \ell < t$---reanchoring that single curve at $(\ell, E(\ell; c), \pi, t)$ reproduces $C$ on $[\ell, U]$ while creating at most two boundary records and preserving slope-label monotonicity \eqref{eq:interleaving} (coalescing if $f_{j+1} = t_{j+1} = \ell$). At any resulting equal-anchor pair $(\ell, v, \pi_j, \ell)$ and $(\ell, w, \pi, t)$ with $t > \ell$, the pure-linear curve is dropped if $w \le v$, the legacy curve is dropped if $(v - w) + (c - \min\{\pi, c\})(t - \ell) \le 0$, and otherwise both are retained in normal form.
\item \emph{Upper-bound and strict-prefix clipping:} For $[0, \ell]$ ($0 \le \ell \le U$), discarding candidates with $f_j > \ell$, shortening at most the last remaining endpoint $t_j > \ell$ to $\ell$, and deleting (or coalescing) that single boundary candidate if it violates \eqref{eq:adjacent-cert} at $\ell$ preserves $C$ on $[0, \ell]$ in normal form. For a strict prefix $u < d_i$, return $\emptyset$ if $d_i = 0$; clip to $[0, d_i - 1]$ if $\mathcal{G} = \mathbb{Z}_{\ge 0}$ and $d_i \ge 1$; and if $\mathcal{G} = \mathbb{R}_{\ge 0}$ and $d_i > 0$, discard candidates with $f_j \ge d_i$, shorten any remaining $t_j > d_i$ to $d_i$, and delete that shortened boundary candidate if it violates \eqref{eq:adjacent-cert} at $d_i$.
\item \emph{Merge truncation at $\xi \in [d_i, Y_i] \cap \mathcal{G}$:} Suppose normal-form families $R$ (representing $O_i$ on $[d_i, Y_i]$ at comparator $\alpha_i$, with $\pi_j \ge \beta_i$) and $X$ (representing $N_i$ on $[d_i, Y_i]$ at comparator $\alpha_i$, with $\pi_j = \beta_i$) satisfy $N_i(x) \le O_i(x)$ for all $x \in [\xi, Y_i] \cap \mathcal{G}$. If $\xi = d_i$, then $\min\{O_i, N_i\} = N_i$ is represented directly by $X$. If $\xi > d_i$, clipping $R$ to the strict prefix $u < \xi$ via Claim~2, lower-clipping $X$ to $[\xi, Y_i]$ via Claim~1, joining the two fragments, and deleting the first node of the right fragment if the seam threshold satisfies $\theta \ge \alpha_i$ produces $\min\{O_i, N_i\}$ on $[d_i, Y_i]$ in normal form.
\end{enumerate}
\end{lemma}

\begin{proof}
In Claim~1, Lemma~\ref{lem:two-neighbors}(2) shows that candidate $j$ dominates all $h < j$ on $[\ell, U]$ and equals a pure regular line of slope $c$ there; keeping its label $\pi_j$ leaves $E_j(x; c) = E_j(\ell; c) + c(x - \ell)$ unchanged because $t - f = 0$, while preserving $\pi_1 \ge \dots \ge \pi_m$. Interleaving ensures at most one successor has $f < \ell < t$, and the equal-anchor rule compares the two affine branches at $\ell$ and $t$, between which their difference is monotone and beyond which it is constant.

In Claim~2, shortening the last surviving endpoint to $\ell$ (or $d_i$) leaves values on $[0, \ell]$ (resp.\ $[0, d_i)$) unchanged; if that shortened candidate $k$ satisfies $E_{k-1}(\ell; c) \le E_k(\ell; c)$ (resp.\ at $d_i$), Lemma~\ref{lem:two-neighbors}(1) implies $E_{k-1}(x; c) \le E_k(x; c)$ across $[f_k, \ell]$ (resp.\ $[f_k, d_i]$), so deleting $k$ preserves the prefix envelope and leaves the unshortened predecessor $k - 1$ satisfying \eqref{eq:adjacent-cert}.

In Claim~3, when $\xi > d_i$, strict-prefix clipping of $R$ retains $O_i$ on $[d_i, \xi) \cap \mathcal{G}$ and only raises curves of $R$ on $[\xi, Y_i]$, where $N_i(x) \le O_i(x)$ already holds. Every retained candidate $u \in R$ has $t_u \le \xi$ (with $f_u < \xi$, and $t_u \le \xi - 1$ on $\mathbb{Z}_{\ge 0}$) and $\pi_u \ge \beta_i$, whereas the first candidate $w$ of clipped $X$ has anchor $f_w = \xi$ and $\pi_w = \beta_i$, so \eqref{eq:interleaving} and strict key ordering $(t_u, f_u) < (t_w, f_w)$ hold across the seam. If $\theta_u \ge \alpha_i$ deletes $w$, any successor $w' = \mathrm{succ}(w)$ in $X$ has anchor $F_{r+1} > \xi$ with strict intercept drop $V_{r+1} < V_r + \beta_i(F_{r+1} - F_r) \le E_u(F_{r+1}; \alpha_i)$ and slope $\beta_i \le \alpha_i$ up to $T_{w'}$, forcing $E_u(T_{w'}; \alpha_i) > E_{w'}(T_{w'}; \alpha_i)$ and hence $\theta_u < \alpha_i$ immediately.
\end{proof}

\section{Balanced-Tree Implementation and Amortized Complexity}
\label{sec:implementation}

We implement the auxiliary recurrence \eqref{eq:regular-op}--\eqref{eq:aux-update} using an augmented height-balanced binary search tree $T$ supporting worst-case $O(\log m)$-time search, insertion, deletion, split, and join on $m$ nodes~\cite{blelloch2016join,brass2008advanced}.

\subsection{Node augmentation and lazy-tag algebra}

Nodes are ordered by $(t_j, f_j)$. Each node stores $(f_j, v_j, \pi_j, t_j)$, a cached successor anchor $\mathrm{succAnchor}_j$, the exact threshold $\theta_j$ of Lemma~\ref{lem:threshold} and anchor secant $R_j = (v_{j+1} - v_j)/(f_{j+1} - f_j)$ (with $R_j = +\infty$ when $f_j = f_{j+1}$, and $\theta_m = R_m = -\infty$ at the rightmost node of a fragment), scalar subtree maxima $\mathrm{maxTheta}$ and $\mathrm{maxR}$, subtree first/last anchors and endpoints, and constant-size lazy tags:
\begin{enumerate}
\item an affine coordinate/value tag $\mathcal{T}(\delta_x, \delta_s, \delta_v)$ acting as $(x, v) \mapsto (x + \delta_x,\, v + \delta_s x + \delta_v)$, composed via
\begin{equation}
  \mathcal{T}(\delta_{x,2}, \delta_{s,2}, \delta_{v,2}) \circ \mathcal{T}(\delta_{x,1}, \delta_{s,1}, \delta_{v,1})
  = \mathcal{T}(\delta_{x,1} + \delta_{x,2},\; \delta_{s,1} + \delta_{s,2},\; \delta_{v,1} + \delta_{v,2} + \delta_{s,2}\delta_{x,1});
  \label{eq:tag-composition}
\end{equation}
\item a legacy-slope tag storing an optional bulk assignment $\pi \leftarrow \beta$ plus an accumulated holding shift $+\Delta_h$ on finite labels;
\item a per-node endpoint-mode bit (`fixed` vs.\ `succSnapshot`, where `succSnapshot` evaluates $t_j = \mathrm{succAnchor}_j$ in $\mathbb{R}_{\ge 0}$ and $\mathrm{succAnchor}_j - 1$ in $\mathbb{Z}_{\ge 0}$) paired with a lazy subtree flag `setSuccSnapshot` that switches all nodes in a subtree to `succSnapshot` mode as tags are pushed down;
\item a threshold-key tag supporting uniform holding shifts $(\theta, R) \mapsto (\theta + \Delta_h, R + \Delta_h)$ and lazy bulk copy $(\theta, R) \mapsto (R, R)$, which commute: $H_{\Delta_h} \circ B = B \circ H_{\Delta_h}$.
\end{enumerate}
Queries $\textsc{ArgMaxTheta}(T)$ and $\textsc{ArgMaxR}(T)$, together with inorder neighbor lookups $\mathrm{pred}(j)$ and $\mathrm{succ}(j)$, descend from the root in $O(\log m)$ time using child subtree aggregates while pushing pending lazy tags along the visited root-to-node paths.

Crucially, `succSnapshot` represents the successor anchor \emph{at the moment of bulk normalization}, not a live pointer to future successors. Before any structural edit mutates an inorder adjacency, the affected root-to-node search paths are exposed (pushing pending lazy tags) and the owner's `succSnapshot` endpoint is materialized into `fixed` mode before its cached $\mathrm{succAnchor}$ is updated. Consequently, deleting a neighbor never retroactively extends an already-truncated legacy branch, and every split, join, or constant-size crop repairs $O(1)$ adjacencies in $O(\log m)$ time.

\subsection{Regular closure, bulk rays, and bracket-guarded merge}

\begin{figure}[tbp]
\noindent\rule{\textwidth}{0.8pt}\par\smallskip
\noindent\textbf{Algorithm 1:} $\textsc{RegularClosure}(T, c)$ --- Regular closure and exact threshold pruning.\par\smallskip
\noindent\rule{\textwidth}{0.4pt}\par\smallskip
\noindent\textbf{Input:} Normal-form tree $T$ with stored comparator $c_{\mathrm{old}}$, new comparator $0 \le c \le c_{\mathrm{old}}$.\\
\textbf{Output:} Pruned normal-form tree $T$ at comparator $c$.
\begin{enumerate}[label=\footnotesize\arabic*:, leftmargin=2.4em, itemsep=2pt, topsep=4pt]
\item Set the global regular comparator of $T$ to $c$.
\item \textbf{while} $T \neq \emptyset$ \textbf{and} $T.\mathrm{maxTheta} \ge c$ \textbf{do}
\item \hspace*{1.2em} Locate $j \leftarrow \textsc{ArgMaxTheta}(T)$ and $k \leftarrow \mathrm{succ}(j)$ by top-down search, exposing the root-to-node paths of $j$, $k$, and their immediate inorder neighbors and freezing any pending `succSnapshot` endpoints on affected predecessors.
\item \hspace*{1.2em} Delete node $k$ from $T$ (keeping the candidate equation of $j$ unchanged) and coalesce if a coincident pure-linear pair arises.
\item \hspace*{1.2em} Recompute $\mathrm{succAnchor}$, $\theta$, and $R$ on the changed adjacency, rebalance, and update subtree aggregates.
\item \textbf{end while}
\item \textbf{return} $T$.
\end{enumerate}
\noindent\rule{\textwidth}{0.8pt}
\end{figure}

\begin{figure}[tbp]
\noindent\rule{\textwidth}{0.8pt}\par\smallskip
\noindent\textbf{Algorithm 2:} $\textsc{DiscountedRays}(X, Q, \beta_i, Y_i)$ --- Bulk ray generation and lazy normalization.\par\smallskip
\noindent\rule{\textwidth}{0.4pt}\par\smallskip
\noindent\textbf{Input:} Prefix tree $X$ representing $O_i$ on $u < d_i$, threshold $Q$, price $\beta_i$, departure bound $Y_i$.\\
\textbf{Output:} Normalized parallel-ray tree $X$ representing $N_i$ on $[0, Y_i]$.
\begin{enumerate}[label=\footnotesize\arabic*:, leftmargin=2.4em, itemsep=2pt, topsep=4pt]
\item If $X = \emptyset$, \textbf{return} $\emptyset$.
\item Apply the lazy shift $\mathcal{T}(Q, 0, Q\beta_i)$ to $X$, mapping $(f_j, v_j) \mapsto (f_j + Q, v_j + Q\beta_i)$, and discard the suffix of $X$ with $f_j > Y_i$.
\item \textbf{while} $X \neq \emptyset$ \textbf{and} $X.\mathrm{maxR} \ge \beta_i$ \textbf{do}
\item \hspace*{1.2em} Locate $j \leftarrow \textsc{ArgMaxR}(X)$ and $k \leftarrow \mathrm{succ}(j)$ by top-down search, exposing paths and freezing affected `succSnapshot` endpoints.
\item \hspace*{1.2em} If $f_j = f_k$, delete whichever of $j, k$ has the larger anchor value $v$ (breaking ties canonically); otherwise delete $k$.
\item \hspace*{1.2em} Repair $\mathrm{succAnchor}$ and $R$ on the updated adjacency, rebalance, and recompute subtree aggregates.
\item \textbf{end while}
\item \textbf{if} $X \neq \emptyset$ \textbf{then}
\item \hspace*{1.2em} Apply the root lazy tags setting $\pi_j \leftarrow \beta_i$ and `setSuccSnapshot` across $X$.
\item \hspace*{1.2em} Expose the rightmost spine of $X$ (pushing `setSuccSnapshot` along it) and set the rightmost node's mode to `fixed` with endpoint $Y_i$.
\item \hspace*{1.2em} Lazily copy active thresholds from secants: $(\theta, \mathrm{maxTheta}) \leftarrow (R, \mathrm{maxR})$.
\item \textbf{end if}
\item \textbf{return} $X$.
\end{enumerate}
\noindent\rule{\textwidth}{0.8pt}
\end{figure}

Algorithms~1 and~2 implement Lemmas~\ref{lem:regular-closure}--\ref{lem:bulk-normalization}. Every nonterminating iteration of either loop deletes a node permanently in $O(\log m)$ time, and the final bulk normalization in Algorithm~2 takes $O(\log m)$ time with zero linear scan.

After splitting $O_i$ into $X$ ($u < d_i$) and $R$ ($u \ge d_i$), constructing $X \leftarrow \textsc{DiscountedRays}(X, Q, \beta_i, Y_i)$, and lower-clipping $X$ to $[d_i, Y_i]$, we merge $R$ and $X$ across the overlap $[L_i, Y_i]$, where $L_i = \max\{Q, d_i\} \le Y_i = d_i + 2Q$. Algorithm~3 gives the backward-sweep merge, and Lemma~\ref{lem:crossover} establishes the $O(\log m)$ bracket-guarded crossover search.

\begin{figure}[tbp]
\noindent\rule{\textwidth}{0.8pt}\par\smallskip
\noindent\textbf{Algorithm 3:} $\textsc{MergeFrontiers}(R, X, L_i, Y_i)$ --- Single-crossing frontier merge.\par\smallskip
\noindent\rule{\textwidth}{0.4pt}\par\smallskip
\noindent\textbf{Input:} Old tree $R$ on $[d_i, Y_i]$, new bulk tree $X$ on $[d_i, Y_i]$, overlap $[L_i, Y_i]$ with $L_i = \max\{Q, d_i\}$.\\
\textbf{Output:} Normal-form tree representing $F_i = \min\{O_i, N_i\}$ on $[d_i, Y_i]$.
\begin{enumerate}[label=\footnotesize\arabic*:, leftmargin=2.4em, itemsep=2pt, topsep=4pt]
\item If $X = \emptyset$ or $L_i > Y_i$, \textbf{return} $R$.
\item If $\textsc{Value}(X, Y_i) > \textsc{Value}(R, Y_i)$, discard $X$ and \textbf{return} $R$.
\item Set right bracket $e \leftarrow Y_i$.
\item \textbf{while} $R \neq \emptyset$ \textbf{do}
\item \hspace*{1.2em} Let $u \leftarrow \textsc{Last}(R)$ and $v \leftarrow \mathrm{pred}(u)$; set $s \leftarrow \max\{L_i, t_v\}$ if $v$ exists, else $s \leftarrow L_i$.
\item \hspace*{1.2em} \textbf{if} $\textsc{Value}(X, s) \le \textsc{Value}(R, s)$ \textbf{then}
\item \hspace*{2.4em} If $f_u < s$ (which forces $s = L_i$), set $z \leftarrow L_i$ and \textbf{break}.
\item \hspace*{2.4em} Freeze affected snapshots, delete $u$ from $R$, repair neighbors, and set $e \leftarrow s$.
\item \hspace*{1.2em} \textbf{else}
\item \hspace*{2.4em} Expose the at-most-two candidate equations of $R$ governing $[s, e)$.
\item \hspace*{2.4em} Set $z \leftarrow \textsc{FirstTakeover}(X, [s, e], \text{exposed old equations})$ and \textbf{break}.
\item \hspace*{1.2em} \textbf{end if}
\item \textbf{end while}
\item If $R = \emptyset$, set $z \leftarrow L_i$.
\item Apply merge truncation at $\xi = z$ via Lemma~\ref{lem:clipping}(3) (returning $X$ if $z = d_i$; otherwise clipping $R$ to $u < z$, lower-clipping $X$ to $[z, Y_i]$, joining the fragments, and pruning at most one seam node if $\theta \ge \alpha_i$), and \textbf{return} the resulting tree.
\end{enumerate}
\noindent\rule{\textwidth}{0.8pt}
\end{figure}

\begin{lemma}[Bracket-guarded crossover search and merge complexity]
\label{lem:crossover}
In Algorithm~3, whenever $\textsc{FirstTakeover}$ is invoked on $[s, e]$, we have $L_i \le s < e \le Y_i$, $\textsc{Value}(X, s) > O_i(s)$, $\textsc{Value}(X, e) \le O_i(e)$, and $O_i(x)$ on $[s, e)$ equals the minimum of at most two exposed candidate equations $(E_{\mathrm{left}}, E_{\mathrm{right}})$ of $R$. Guarding the binary search on $X$ so that anchors $F_w \le s$ record $a \leftarrow w$ and branch right without evaluating \eqref{eq:guarded-predicate}, anchors $F_w \ge e$ branch left without evaluating \eqref{eq:guarded-predicate}, and the predicate
\begin{equation}
  V_w \le \min\bigl\{E_{\mathrm{left}}(F_w; \alpha_i),\; E_{\mathrm{right}}(F_w; \alpha_i)\bigr\}
  \label{eq:guarded-predicate}
\end{equation}
is evaluated only for anchors $F_w \in (s, e)$ (recording $a \leftarrow w$ and branching right on failure, or recording $b \leftarrow w$ and branching left on success) locates the exact takeover coordinate $z \in (s, e]$ in $O(\log m)$ time without nested tree searches. Consequently, a merge deleting $k$ nodes from $R$ runs in $O((k + 1)\log m)$ time, creates $O(1)$ boundary nodes, and outputs the normal form of $F_i$ on $[d_i, Y_i]$.
\end{lemma}

\begin{proof}
Each deleted node $u$ satisfies $f_u \ge s$ and $N_i(s) \le O_i(s)$, so removing $u$ leaves $\textsc{Value}(R, x) = O_i(x)$ unchanged for all $x < s$ and cannot lower $\textsc{Value}(R, s)$ at $e \leftarrow s$, preserving $\textsc{Value}(X, e) \le \textsc{Value}(R, e)$. Thus entering the `else` branch forces $s < e$, with no interior endpoint of $R$ inside $(s, e)$, so Lemma~\ref{lem:two-neighbors}(3) expresses $O_i(x)$ on $[s, e)$ via at most two exposed equations $(E_{\mathrm{left}}, E_{\mathrm{right}})$ of $R$.

Restricting \eqref{eq:guarded-predicate} to $F_w \in (s, e)$ guarantees that $\min\{E_{\mathrm{left}}(F_w; \alpha_i), E_{\mathrm{right}}(F_w; \alpha_i)\} = O_i(F_w)$ is never evaluated outside the valid domain of the exposed pair (avoiding spurious $+\infty$ values at anchors of $X$ below $f_{\mathrm{left}} \le s$), while recording $a \leftarrow w$ whenever $F_w \le s$ is valid because $\textsc{Value}(X, s) > O_i(s)$ rules out takeover at or before $s$. By Theorem~\ref{thm:single-crossing}, \eqref{eq:guarded-predicate} is monotone across $F_w \in (s, e)$. Initializing $a \leftarrow \mathrm{nil}$ and sentinel $F_b \leftarrow e$, the $O(\log m)$ search finds the last anchor $F_a < e$ not yet winning at its anchor (if any exists) and the first anchor $F_b > s$ winning at its anchor (or sentinel $e$). On $(\max\{s, F_a\}, \min\{e, F_b\})$ (taking $\max\{s, F_a\} = s$ when $a = \mathrm{nil}$), either $a = \mathrm{nil}$ ($N_i = +\infty$ there, so takeover occurs at the jump $z = F_b$), or $N_i(x) = V_a + \beta_i(x - F_a)$ is a single line of slope $\beta_i$ compared against the constant-size piecewise-affine curve $\min\{E_{\mathrm{left}}, E_{\mathrm{right}}\}$ (solving $O(1)$ affine inequalities, or rounding up in $\mathbb{Z}_{\ge 0}$, and defaulting to $z = F_b$ if no crossing precedes $F_b$). Truncation and seam repair follow Lemma~\ref{lem:clipping}(3).
\end{proof}

\subsection{Online per-period algorithm and proof of Theorem~\ref{thm:main}}

\begin{figure}[tbp]
\noindent\rule{\textwidth}{0.8pt}\par\smallskip
\noindent\textbf{Algorithm 4:} Online $O(n\log n)$ Optimal-Value Algorithm for Arbitrary Uncapacitated Prices.\par\smallskip
\noindent\rule{\textwidth}{0.4pt}\par\smallskip
\noindent\textbf{Input:} Horizon $n$, threshold $Q > 0$, sequential stream $(d_i, p_{1,i}, p_{2,i}, h_i)$ for $i = 1, \dots, n$.\\
\textbf{Output:} Free-terminal optimum $\operatorname{OPT}$ and zero-terminal prefix optima $(Z_1, \dots, Z_n)$.
\begin{enumerate}[label=\footnotesize\arabic*:, leftmargin=2.4em, itemsep=2pt, topsep=4pt]
\item Initialize $T$ with the single candidate $(f_1 = 0, v_1 = 0, \pi_1 = +\infty, t_1 = 0)$ on $[0, 0]$; set $K \leftarrow 0$.
\item \textbf{for} $i = 1, \dots, n$ \textbf{do}
\item \hspace*{1.2em} If $i = 1$, set $\alpha_1 \leftarrow p_{1,1}$, $\beta_1 \leftarrow p_{2,1}$; else set $\alpha_i \leftarrow \min\{\alpha_{i-1} + h_{i-1}, p_{1,i}\}$ and $\beta_i \leftarrow \min\{\beta_{i-1} + h_{i-1}, p_{2,i}\}$.
\item \hspace*{1.2em} Set $Y_i \leftarrow d_i + 2Q$; extend the target domain of $T$ to $[0, Y_i]$ at comparator $\alpha_i$ (if $i = 1$, stored comparator starts at $\alpha_1$).
\item \hspace*{1.2em} $T \leftarrow \textsc{RegularClosure}(T, \alpha_i)$.
\item \hspace*{1.2em} $(X, R) \leftarrow \textsc{SplitFrontier}(T,\; u < d_i,\; u \ge d_i)$ via Lemma~\ref{lem:clipping}(1)--(2).
\item \hspace*{1.2em} $X \leftarrow \textsc{DiscountedRays}(X, Q, \beta_i, Y_i)$; lower-clip $X$ to targets $y \ge d_i$.
\item \hspace*{1.2em} $T \leftarrow \textsc{MergeFrontiers}(R, X, \max\{Q, d_i\}, Y_i)$; normalize $T$ on $[d_i, Y_i]$.
\item \hspace*{1.2em} Record zero-terminal prefix optimum $Z_i \leftarrow \textsc{Value}(T, d_i)$.
\item \hspace*{1.2em} Apply lazy travel/holding tag $\mathcal{T}(-d_i, h_i, -h_i d_i)$, increment finite labels $\pi_j$, comparator $\alpha_i$, and thresholds $(\theta_j, R_j)$ by $+h_i$, and set arrival domain $K \leftarrow 2Q$.
\item \textbf{end for}
\item Push all remaining lazy tags in a single $O(n)$ traversal of $T$ and compute $\operatorname{OPT} \leftarrow \min_{j \in T} v_j$.
\item \textbf{return} $\operatorname{OPT}$ and $(Z_1, \dots, Z_n)$.
\end{enumerate}
\noindent\rule{\textwidth}{0.8pt}
\end{figure}

\begin{proof}[Proof of Theorem~\ref{thm:main}]
We first verify correctness by induction on $i$. At $i = 1$, $T$ represents $A_1(0) = 0$ on $[0, 0]$. Suppose $T$ represents $A_i$ in normal form on $[0, K_i]$ at entry to period $i$, with stored comparator $\alpha_{i-1} + h_{i-1}$ (for $i \ge 2$) and all finite legacy labels satisfying $\pi_j \ge \beta_{i-1} + h_{i-1}$. By \eqref{eq:nonspeculative}, $\alpha_i \le \alpha_{i-1} + h_{i-1}$ and $\beta_i \le \beta_{i-1} + h_{i-1} \le \pi_j$, so Lemma~\ref{lem:regular-closure} and Algorithm~1 produce $O_i$ on $[0, Y_i]$ (note $K_i = 2Q \le d_i + 2Q = Y_i$, so no incoming upper crop is even needed). By Lemma~\ref{lem:closure} ($\mathcal{D}_{\beta_i, d_i}\mathcal{O}_{\alpha_i} A_i = \mathcal{D}_{\beta_i, d_i} A_i$) and Lemma~\ref{lem:bulk-normalization}, Step~7 produces $N_i$ restricted to $[d_i, Y_i]$. By Lemma~\ref{lem:crossover}, Step~8 produces $F_i = \min\{O_i, N_i\}$ on $[d_i, Y_i]$, so Step~9 records $Z_i = F_i(d_i) = A_{i+1}(0) = \operatorname{OPT}_{i,0}$ by Theorem~\ref{thm:aux-exactness}. Step~10 maps departure coordinate $y = d_i + z \in [d_i, d_i + 2Q]$ to arrival coordinate $z \in [0, 2Q]$ and adds $h_i(y - d_i) = h_i z$, yielding $A_{i+1}(z) = F_i(d_i + z) + h_i z$ with comparator $\alpha_i + h_i$ and finite labels $\ge \beta_i + h_i$. After period $n$, every candidate slope in $A_{n+1}$ is at least $\beta_n + h_n \ge 0$, so each candidate attains its minimum at its anchor $f_j \in [0, 2Q]$, giving $\min_{0 \le z \le 2Q} A_{n+1}(z) = \min_{j \in T} v_j = \operatorname{OPT}$ by Theorem~\ref{thm:aux-exactness}.

For time and space complexity, $T$ starts with $1$ node, and each period $i$ performs $O(1)$ splits, clips, and joins that create at most a constant number of boundary nodes; bulk updates mutate existing nodes lazily without copying. Hence at most $O(n)$ candidate nodes are ever created, bounding $|T| = O(n)$ at all times and total deletions across all invocations of Algorithms~1--3 by $O(n)$. Charging each nonterminal while-loop iteration to its permanent node deletion at $O(\log n)$ per deletion, and adding $O(\log n)$ per period for constant-count splits, point queries, and the guarded crossover search, yields $O(n\log n)$ total time. Each node stores $O(1)$ fields and no historical tree versions are retained, so working space is $O(n)$ (and strictly $O(n)$ even if only $\operatorname{OPT}$ is streamed, with $O(1)$ scalar state outside $T$). Note that Steps~2--9 require no knowledge of $n$ or future periods.
\end{proof}

\section{Two-Stage Recovery of an Optimal Physical Schedule}
\label{sec:recovery}

Because an auxiliary candidate's stored anchor cost $v_j$ need not equal the lower envelope at $f_j$, schedule reconstruction cannot simply store arbitrary parent pointers per node; furthermore, virtual orders in $(\alpha_i, \beta_i)$ must be mapped back to physical purchasing periods. We solve this in two clean stages.

\subsection{Persistent frontier snapshots and virtual predecessor formulas}

Implement $T$ via standard path-copying persistence over immutable child links, scalar subtree extrema ($\mathrm{maxTheta}$, $\mathrm{maxR}$, and anchor/endpoint bounds), and cached scalar $\mathrm{succAnchor}$ fields~\cite{blelloch2016join,brass2008advanced}. Because $\textsc{ArgMaxTheta}$, $\textsc{ArgMaxR}$, neighbor lookups, splits, and joins all operate top-down from the current root using child aggregates, each search-path edit copies $O(\log n)$ nodes, pushes lazy tags onto the copied path nodes, and shares unmodified subtrees without parent or mutable thread pointers. Alongside the $O(n)$ scalar arrays $(\alpha_i, \beta_i, s(i), r(i))$, save two root pointers in each period $i \in \{1, \dots, n\}$:
\begin{enumerate}
\item $T^{\mathrm{reg}}_i$: the root of $T$ immediately after Step~5 ($\textsc{RegularClosure}(T, \alpha_i)$), representing $O_i$ on $[0, Y_i]$ \emph{before} splitting at $d_i$;
\item $T^{\mathrm{disc}}_i$: the root of $X$ immediately after $\textsc{DiscountedRays}(X, Q, \beta_i, Y_i)$ in Step~7, representing $N_i$ on $[0, Y_i]$ \emph{before} lower-clipping at $d_i$ or merge reanchoring, so every surviving bulk record $(F_j, V_j)$ satisfies $F_j = f_j + Q$ and $V_j = v_j + Q\beta_i$ for an original incoming anchor $f_j < d_i$.
\end{enumerate}

\begin{lemma}[Exact virtual predecessors]
\label{lem:virtual-predecessors}
Fix period $i$ and target departure inventory $y \in [d_i, Y_i] \cap \mathcal{G}$.
\begin{enumerate}
\item If candidate $j = (f_j, v_j, \pi_j, t_j)$ in $T^{\mathrm{reg}}_i$ attains $O_i(y)$, then
\begin{equation}
  u =
  \begin{cases}
    f_j, & \pi_j \ge \alpha_i, \\
    \min\{y, t_j\}, & \pi_j < \alpha_i,
  \end{cases}
  \label{eq:reg-pred}
\end{equation}
satisfies $u \in [0, K_i] \cap \mathcal{G}$, $u \le y$, and $O_i(y) = A_i(u) + \alpha_i(y - u)$.
\item If $N_i(y) < +\infty$ and $j$ is the last record in $T^{\mathrm{disc}}_i$ with $F_j \le y$, then $u = F_j - Q$ satisfies $u \in [0, K_i] \cap \mathcal{G}$, $u < d_i$, $y - u \ge Q$, and $N_i(y) = A_i(u) + \beta_i(y - u)$.
\end{enumerate}
\end{lemma}

\begin{proof}
In Claim~1, all anchors and endpoints in $T^{\mathrm{reg}}_i$ originate from $A_i$ on $[0, K_i]$, so $u \in [0, K_i] \cap \mathcal{G}$. By the proof of Lemma~\ref{lem:regular-closure}, $E_j(y; \alpha_i) = E_j^{\mathrm{in}}(u) + \alpha_i(y - u)$, where $E_j^{\mathrm{in}}$ is candidate $j$'s curve in $A_i$. Since $A_i(u) \le E_j^{\mathrm{in}}(u)$, a strict inequality $A_i(u) < E_j^{\mathrm{in}}(u)$ would imply $O_i(y) \le A_i(u) + \alpha_i(y - u) < E_j(y; \alpha_i) = O_i(y)$, a contradiction (ruling out, in particular, picking an unattained left-limit endpoint at a downward jump). In Claim~2, strictly decreasing intercepts $\eta_j = V_j - \beta_i F_j$ imply $N_i(y) = V_j + \beta_i(y - F_j) = v_j + \beta_i(y - u)$ at $u = F_j - Q < d_i$; since $A_i(u) \le v_j$ and $A_i(u) < v_j$ would strictly lower $N_i(y)$, we have $A_i(u) = v_j$.
\end{proof}

\subsection{Backward virtual backtracking and forward physical routing}

\begin{proof}[Proof of Theorem~\ref{thm:recovery-main}]
Run the persistent version of Algorithm~4 up to the desired horizon $m = n$ (setting $z^\star = f_{j^\star}$, where $j^\star \in T$ minimizes $v_j$ and breaks ties toward the smallest anchor $f_j$, so $z^\star < Q$ by Theorem~\ref{thm:aux-exactness}) or $m = k$ (setting $z^\star = 0$ for prescribed zero terminal stock). Because Lemma~\ref{lem:crossover} and Theorem~\ref{thm:main} bound the total number of elementary tree searches, splits, joins, and deletions by $O(n)$, path-copying persistence creates $O(n\log n)$ node records in $O(n\log n)$ total forward time and space.

\emph{Stage 1 (Virtual backtracking):} Starting from $\hat{I}_m = z^\star$, iterate $i = m, m-1, \dots, 1$. Set $y = d_i + \hat{I}_i$ and evaluate $O_i(y)$ on $T^{\mathrm{reg}}_i$ and $N_i(y)$ on $T^{\mathrm{disc}}_i$ in $O(\log n)$ time (carrying lazy parameters top-down without mutating saved nodes). Select whichever branch attains $F_i(y) = \min\{O_i(y), N_i(y)\}$, compute the predecessor $u$ via Lemma~\ref{lem:virtual-predecessors}, and set $\hat{q}_i = y - u$ and $\hat{I}_{i-1} = u$. Summing $A_{i+1}(\hat{I}_i) = A_i(\hat{I}_{i-1}) + \kappa_i + h_i \hat{I}_i$ from $i = 1$ to $m$ (with $\hat{I}_0 = 0$) shows that $(\hat{q}_1, \dots, \hat{q}_m)$ is a feasible virtual schedule satisfying $\widehat{J}_m(\hat{q}, \hat{I}) \le \sum_{i=1}^m (\kappa_i + h_i \hat{I}_i) = \operatorname{OPT}_m$ (or $\operatorname{OPT}_{m,0}$), because $\widehat{P}_i(\hat{q}_i) \le \alpha_i \hat{q}_i$ on regular steps and $\widehat{P}_i(\hat{q}_i) = \beta_i \hat{q}_i$ on discounted steps ($\hat{q}_i \ge Q$). Stage~1 performs $m$ backward steps of $O(\log n)$ time each.

\emph{Stage 2 (Retroactive origin aggregation):} Apply \eqref{eq:origin-map}--\eqref{eq:physical-aggregation} of Theorem~\ref{thm:effective-equivalence} in $O(m)$ time using the saved origin indices $(s(i), r(i))$, setting $q_t = \sum_{i:\, \hat{q}_i > 0,\, \tau(i) = t} \hat{q}_i$ for $t = 1, \dots, m$. By Theorem~\ref{thm:effective-equivalence}, $(q_1, \dots, q_m) \in \mathcal{G}^m$ is physically feasible, satisfies $I_m = \hat{I}_m$ (zero when $z^\star = 0$), and achieves $J_m(q, I) \le \widehat{J}_m(\hat{q}, \hat{I}) \le \operatorname{OPT}_m$ (resp.\ $\operatorname{OPT}_{m,0}$), so it is an exact physical optimum. Total time and space are $O(n\log n)$.
\end{proof}

\section{Worked Examples}
\label{sec:examples}

\subsection{Speculative forward buying and virtual-to-physical routing}

Let $Q = 5$, $n = 3$, holding rates $(h_1, h_2, h_3) = (1, 1, 0)$, and consider the price-spike instance:
\[
\begin{array}{c|rrrr}
  i & d_i & p_{1,i} & p_{2,i} & h_i \\ \hline
  1 & 0 & 6 & 2 & 1 \\
  2 & 12 & 25 & 15 & 1 \\
  3 & 2 & 3 & 1 & 0
\end{array}
\]
In physical coordinates, period~1 has zero demand ($d_1 = 0 = I_0$), yet the unique optimal zero-terminal schedule is $q = (12, 0, 2)$ with physical inventories $I = (0, 12, 0, 0)$ and total cost $P_1(12) + h_1(12) + P_2(0) + h_2(0) + P_3(2) = 24 + 12 + 0 + 0 + 6 = 42$, whereas the unique free-terminal optimum is $41$, attained by $q = (12, 0, 5)$ ending with $I_3 = 3 < Q$ at cost $24 + 12 + 5 = 41$. Notice that in both optimal physical schedules:
\begin{itemize}
\item $q_1 = 12 > 0$ despite $I_0 = 0 \ge d_1 = 0$, violating physical zero-inventory ordering;
\item $I_1 = 12 > 2Q = 10$, violating the $2Q$ bound on physical stock (and scaling $d_2$ makes $I_1$ arbitrarily large).
\end{itemize}

Now trace Algorithm~4 and Theorem~\ref{thm:recovery-main}. The online effective prices and origin pointers are:
\begin{align*}
  i = 1:\quad & \alpha_1 = 6\ (s(1)=1), && \beta_1 = 2\ (r(1)=1); \\
  i = 2:\quad & \alpha_2 = \min\{6+1, 25\} = 7\ (s(2)=1), && \beta_2 = \min\{2+1, 15\} = 3\ (r(2)=1); \\
  i = 3:\quad & \alpha_3 = \min\{7+1, 3\} = 3\ (s(3)=3), && \beta_3 = \min\{3+1, 1\} = 1\ (r(3)=3).
\end{align*}
At $i = 1$, $d_1 = 0$ forbids any auxiliary discounted ray ($u < 0$ is empty), so $F_1(y) = O_1(y) = 6y$ on $[0, 10]$, $Z_1 = F_1(0) = 0$, and $A_2(z) = 7z$ on $[0, 10]$. At $i = 2$ ($d_2 = 12$, $Y_2 = 22$), anchor $f_1 = 0 < 12$ generates the discounted ray $N_2(y) = 3y$ on $[5, 22]$, which dominates $O_2(y) = 7y$ throughout $[12, 22]$, giving $Z_2 = F_2(12) = 36$ and $A_3(z) = F_2(12 + z) + z = 36 + 4z$ on $[0, 10]$. At $i = 3$ ($d_3 = 2$, $Y_3 = 12$), lowering the regular comparator from $4$ to $\alpha_3 = 3$ yields $O_3(y) = 36 + 3y$ on $[0, 12]$, while anchor $f_1 = 0 < 2$ generates $N_3(y) = 36 + y$ on $[5, 12]$. Thus $F_3(y) = 36 + 3y$ on $[2, 5)$ and $F_3(y) = 36 + y$ on $[5, 12]$, recording $Z_3 = F_3(2) = 42 = \operatorname{OPT}_{3,0}$. Shifting by $d_3 = 2$ with $h_3 = 0$ leaves two candidates in $A_4$ on $[0, 10]$ with anchor values $v_1 = 42$ (at $f_1 = 0$) and $v_2 = 41$ (at $f_2 = 3$), yielding $\operatorname{OPT} = \min\{42, 41\} = 41$.

For free-terminal recovery ($z^\star = 3$): Period~3 ($y = 5$) selects $N_3(5) = 41$ with $u = 0$ and $\hat{q}_3 = 5 \ge Q$; Period~2 ($z = 0, y = 12$) selects $N_2(12) = 36$ with $u = 0$ and $\hat{q}_2 = 12 \ge Q$; Period~1 ($z = 0, y = 0$) gives $\hat{q}_1 = 0$. Virtual stock never exceeds $z^\star = 3 < 2Q$. Stage~2 then maps $\hat{q}_2 = 12$ to $\tau(2) = r(2) = 1$ and $\hat{q}_3 = 5$ to $\tau(3) = r(3) = 3$, recovering the exact speculative physical plan $q = (12, 0, 5)$ of cost $41$ (and starting from $z^\star = 0$ analogously yields $\hat{q} = (0, 12, 2)$ and physical plan $q = (12, 0, 2)$ of cost $42$).

\subsection{Positive-state discrepancy and three-regime thresholds}

Two further subtleties are shared with the monotone-price auxiliary framework~\cite{papadopoulos2026monotone} and explain our representation choices:
\begin{enumerate}
\item \emph{Intermediate positive states need not match unrestricted costs:} For $Q = 2$, $h = (0, 0)$, $d = (1, 1)$, $p_{1,1} = p_{1,2} = 5$, and $(p_{2,1}, p_{2,2}) = (2, 1)$, reaching terminal stock $I_2 = 2$ has true cost $6$ via $q = (2, 2)$ (where $I_1 = 1 = d_2$), whereas $A_2(z) = 5(1 + z)$ on $[0, 1) \cap \mathcal{G}$ and $A_2(z) = 2(1 + z)$ on $[1, 4] \cap \mathcal{G}$ give $O_2(3) = A_2(3) = 8$ and $N_2(3) = A_2(0) + 3 = 8$ (because the auxiliary recurrence forbids a discounted order from $u = 1 = d_2$), assigning $A_3(2) = 8 > 6$. Nevertheless, both models have $\operatorname{OPT} = 4$ (at $I_2 = 0$ via $q = (2, 0)$) and $\operatorname{OPT}_{2,0} = 4$, consistent with Theorem~\ref{thm:aux-exactness}.
\item \emph{Anchor secants underestimate domination when a legacy branch exists:} Suppose at entry to period~3 (at stored comparator $c_{\mathrm{old}} = 6$) a pruned frontier has left candidate $(f_L, v_L, \pi_L, t_L) = (0, 10, 2, 1)$ and right candidate $(f_R, v_R, \pi_R, t_R) = (2, 17, 1, 2)$. The naive anchor secant is $(17 - 10)/(2 - 0) = 3.5$, yet for any comparator $c \ge 2$ the exact difference in \eqref{eq:three-regimes} is $D(c) = 10 + 2 + c - 17 = c - 5$, giving $\theta = 5$. At $\alpha_3 = 4$, $E_L(2; 4) = 16 < 17 = E_R(2; 4)$, which $\theta = 5$ detects and the secant $3.5$ misses.
\end{enumerate}
\paragraph{Use of AI-Assisted Technologies:} 
Generative AI tools were utilized to enhance clarity, grammar, and overall readability. All core ideas, analyses, and conclusions remain the original work of the author, who has thoroughly reviewed and verified the final text.
\section{Concluding Remarks}
\label{sec:conclusion}

We have shown that uncapacitated single-item lot sizing with a stationary one-breakpoint all-units quantity discount, zero setup cost, nonnegative linear holding costs, and arbitrary time-varying prices is solvable in $O(n\log n)$ time and $O(n)$ working space, matching the classical $O(n\log n)$ complexity of uncapacitated Wagner--Whitin lot sizing~\cite{aggarwal1993improved,wagelmans1992economic}. By reducing arbitrary prices online to non-speculative effective prices $(\alpha_i, \beta_i)$ and reconstructing physical orders via origin pointers $(s(i), r(i))$, the algorithm avoids maintaining explicit speculative inventory states above $2Q$ while remaining exact for both continuous and integral domains.

Two boundaries of the present result merit emphasis. First, the retroactive aggregation in Theorem~\ref{thm:effective-equivalence} relies on unlimited storage capacity ($B_i = +\infty$) because routing later virtual orders back to an earlier origin period $t = \tau(i)$ raises intermediate physical inventories $I_m$ for $t \le m < i$ as shown in \eqref{eq:inventory-lift}. Notice, however, that our proof of Theorems~\ref{thm:normal-form}--\ref{thm:aux-exactness} and Section~\ref{sec:implementation} holds directly whenever raw prices already satisfy the non-speculative conditions $p_{1,i+1} \le p_{1,i} + h_i$ and $p_{2,i+1} \le p_{2,i} + h_i$ (even under arbitrary post-order capacities $B_i \ge d_i$ by setting $K_{i+1} = \min\{2Q, B_i - d_i\}$, $Y_i = \min\{B_i, d_i + 2Q\}$, and upper-clipping $A_i$ to $[0, \min\{K_i, Y_i\}]$ at entry to period $i$, since no origin rerouting is needed when $\tau(i) = i$). Extending near-linear algorithms to capacitated instances that combine speculative price spikes ($p_{2,i+1} > p_{2,i} + h_i$) with fluctuating hard storage bottlenecks remains a distinct structural problem. Second, under an \emph{incremental} quantity discount (where $p_{2,i} \le p_{1,i}$ applies only to units ordered above $Q$), the procurement cost is continuous and piecewise-linear \emph{concave} rather than discontinuous; investigating whether $\Omega(n\log n)$ is a provable algebraic decision-tree lower bound for uncapacitated one-breakpoint all-units lot sizing under arbitrary prices is an appealing question for future work.

\end{document}